\documentclass[a4paper,11pt]{article}
\pdfoutput=1 

\usepackage{jcappub} 

\usepackage[T1]{fontenc} 

\title{\boldmath 20 GeV halo-like excess of the Galactic diffuse emission
and implications for dark matter annihilation}

\author[a,b]{Tomonori Totani}

\affiliation[a]{Department of Astronomy, School of Science, The University of Tokyo, Bunkyo-ku, Tokyo 113-0033, Japan}
\affiliation[b]{Research Center for the Early Universe, School of Science, The University of Tokyo, Bunkyo-ku, Tokyo 113-0033, Japan}

\emailAdd{totani@astron.s.u-tokyo.ac.jp}

\abstract{
Fifteen years of the {\it Fermi} Large Area Telescope (LAT) data in the halo region of the Milky Way (MW) are analyzed to search for gamma rays from dark matter annihilation.
Gamma-ray maps within the region of interest ($|l| \le 60^\circ$, $10^\circ \le |b| \le 60^\circ$) are modeled using point sources, the GALPROP models of cosmic-ray interactions, isotropic background, and templates of Loop I and the Fermi bubbles, and then the presence of a halo-like component is further examined.
A statistically significant halo-like excess is found with a spectral peak around 20 GeV, while its flux is consistent with zero below 2 GeV and above 200 GeV.
Examination of the fit residual maps indicates that a spherically symmetric halo component fits the map data well.
The radial profile agrees with annihilation by the smooth NFW density profile, and may be slightly shallower than this, especially in the central region.
Various systematic uncertainties are investigated, but the 20 GeV peak remains significant.
In particular, the halo excess with a similar spectrum is detected even relative to the LAT standard background model, which contains non-template patches adjusted to match the observed map. 
The halo excess spectrum can be fitted by annihilation with a particle mass $m_\chi \sim$ 0.5--0.8 TeV and cross section $\langle \sigma \upsilon \rangle \sim$ (5--8)$\times 10^{-25} \ \rm cm^3 \, s^{-1}$ for the $b\bar{b}$ channel.
This cross section is larger than the upper limits from dwarf galaxies and the canonical thermal relic value, but considering various uncertainties, especially the density profile of the MW halo, the dark matter interpretation of the 20 GeV ``Fermi halo'' remains feasible.
The prospects for verification through future observations are briefly discussed.
}

\begin{document}
\maketitle
\flushbottom

\section{Introduction}  
\label{sec:intro}

Even though nearly a century has passed since Fritz Zwicky inferred the existence of dark 
matter from the motion of galaxies in the Coma cluster \cite{Zwicky1933}, dark matter remains one of the 
greatest mysteries in astrophysics, cosmology, and fundamental 
physics \cite{Bertone2018,Arbey2021,Cirelli2024}. 
Weakly interacting massive particles (WIMPs) in a mass range from 1 GeV to 100 TeV near the electroweak scale have long been considered one of the leading candidates, because such particles are naturally predicted in some particle physics theories beyond the Standard Model, and because the thermal relic amount predicted from the annihilation cross section expected for WIMPs is close to the observed dark matter density  \cite{Arbey2021,Cirelli2024,Arcadi2018,Roszkowski2018}. 
The existence of WIMPs can be probed with direct searches by underground experiments, 
indirect searches using annihilation products (e.g., gamma-rays and neutrinos), and by particle accelerator experiments.

Regarding indirect searches by gamma rays, the sensitivities of the {\it Fermi} Large Area Telescope ({\it Fermi} LAT) and ground-based air Cherenkov gamma-ray telescopes have reached a level where the intensity expected from the canonical annihilation cross section can be detected. 
When WIMPs directly annihilate into two photons, line (monoenergetic) gamma-ray emission is easy to distinguish, but the branching ratio into line emission is highly uncertain.
In the more mainstream annihilation mode, the Standard Model particles such as $b\bar{b}$ 
or $W^+W^-$ are first generated, followed by a cascade that produces gamma rays with a continuum energy spectrum.  
The spectrum still shows a broad peak around a specific gamma-ray energy related to the WIMP mass, which contrasts with power-law spectra of many astrophysical objects.
Numerous explorations have been conducted toward various astronomical objects or regions in search of such signals, including the Milky Way (MW) Galactic center (GC) \cite{Ackermann2017-GCE,Murgia2020,Abdalla2022}, the Large and Small Magellanic Clouds (LMC and SMC) \cite{Buckley2015,Caputo2016}, dwarf galaxies \cite{Ackermann2015-dSph,Albert2017,Acciari2020,McDaniel2024} and dark subhalos \cite{Ackermann2012-subhalo,Coronado2022} in the MW halo, the Andromeda galaxy \cite{Albert2018,Karwin2021}, galaxy groups and clusters \cite{Arlen2012,Ackermann2015-cl,Lisanti2018,DiMauro2023}, and the extragalactic background radiation \cite{Fermi-LAT2015-DM-EGB,Fornasa2015,Fornasa2016,Cholis2024}. 

This study will search for annihilation gamma rays from the MW halo, avoiding the GC and disk regions. Although the annihilation radiation spreading across the MW halo is not as strong as that near the GC, it is possible to avoid the influence of gamma rays originating from various objects and cosmic rays around the GC and the disk, thereby potentially increasing the effective signal-to-noise ratio.
Looking back on past searches towards the MW halo \cite{Ackermann2012-MWhalo,Mazziotta2012,GomezVargas2013,Tavakoli2014,Ackermann2015-line,Chang2018,Albert2023}, it seems that interest in this region for a dark matter search has waned compared to other objects or regions since the early 2010s. This is probably because the so-called GC GeV excess with a sharp morphological peak at the GC, which may be a result of dark matter annihilation, has been found around a photon energy of a few GeV \cite{Goodenough2009,Vitale2009,Ackermann2017-GCE,Murgia2020}.
Another possible reason is that halo-scale (up to $\sim$50 deg from the GC) diffuse radiation structures, such as the Fermi bubbles \cite{Dobler2010,Su2010,Ackermann2014-FB}, have been found. 
They are unlikely to originate from dark matter and are confusing when searching for an annihilation signal in the MW halo region. 

However, the GC GeV excess can also be explained by astrophysical phenomena rather than dark matter, such as millisecond pulsars \cite{Murgia2020}. 
Then the true dark matter signal may appear in other photon energy ranges. 
To explain the GC excess with dark matter annihilation, the radial density profile near the GC must be $\rho \propto r^\gamma$ with $\gamma = $1.1--1.3, which is more centrally concentrated than the standard Navarro-Frenk-White (NFW) \cite{Navarro1997} profile ($\gamma = 1$ in the central region).
If the true density profile is NFW or less centrally concentrated, dark matter signals may be more easily detected in the halo region than at the GC.
In 2012, the {\it Fermi} LAT team analyzed the MW halo region in search of dark matter using two years of the LAT data \cite{Ackermann2012-MWhalo}, which is much smaller than the amount accumulated to date. 
From that time until now, understanding and analysis methods for diffuse radiation in the halo region, such as the Fermi bubbles and Loop I, have also matured.
For these reasons, it would be worthwhile to investigate the presence or absence of a halo-like radiation in addition to the known components based on the latest LAT data.
This study reports the results of a dark matter search using 15 years of the LAT data.

The structure of this paper is as follows. 
Section \ref{sec:methods} describes the LAT data used in this work, the region of interest in the analysis, the likelihood analysis method, and various model map templates considered. 
The results of the search for a halo-like excess are presented in section \ref{sec:results}.
Section \ref{sec:DM} discusses the interpretation of the results in terms of dark matter annihilation, and section \ref{sec:summary} presents an overview and some discussion.

\section{Analysis methods}
\label{sec:methods}

\subsection{The LAT data and the region of interest}

This work is based on the Pass 8 UltraClean class of
the 15-year public {\it Fermi} LAT data from 2008 August 4 to 2023 August 2. 
In addition, a zenith angle cut of $\theta < 100^\circ$ is adopted to reduce the influence of radiation from the Earth's atmosphere.
To avoid deterioration of angular resolution in the low photon energy range, this study analyzes only the range above 1 GeV. 
The angular resolution (68\% containment angles) of the LAT is 0.85$^\circ$, 0.16$^\circ$, and 0.10$^\circ$ at 1, 10, and 100 GeV, respectively\footnote{\url{https://www.slac.stanford.edu/exp/glast/groups/canda/lat_Performance.htm}} \cite{Ajello2021}.
The data are divided into 13 energy bins, equally spaced in logarithmic space, with the minimum and maximum energy bin centers at 1.51 and 814 GeV, respectively. 

The region of interest (ROI) in this study is set to $|l| \le 60^\circ$ and $|b| \le 60^\circ$ around the GC, and the disk area of $|b| < 10^\circ$ is excluded from the dark matter search. 
Photon count maps and exposure maps with pixel scales of $\Delta l_p = \Delta b_p = 0.125^\circ$ in Cartesian coordinates are created from the public {\it Fermi} LAT data and tools\footnote{\url{https://fermi.gsfc.nasa.gov/ssc/data/}}, resulting in 960$\times$960 pixels in the ROI including the disk region.

\subsection{Likelihood analyses}
The maximum likelihood fitting is performed independently for each photon energy bin. 
In many studies that fit a model map to the LAT data, the photon counts and the model expectations are compared in each pixel. 
Since the halo-like component is expected to be much smoother than the pixel scale of 0.125$^\circ$,  calculating the likelihood at a coarser resolution allows computation time to be reduced without compromising dark matter search.
Therefore, we divide the ROI into $12 \times 12$ cells with a width of $\Delta l_c = \Delta b_c = 10^\circ$ and calculate the likelihood based on photon counts and expectations in these cells. 
Some components of the Galactic diffuse emission, such as those from interstellar gas, possess pixel-scale fine structures, which may help separate them from others by pixel-scale likelihood calculation.
On the other hand, smoothing at coarse resolutions may reduce the impact of pixel-scale mismatches between a model map and data on the likelihood estimate. 

Let $C_{\rm obs, \it i}$ be the observed photon count in the $i$-th cell and $C_{\exp, i}$ be the number of photons expected from the model, and then the likelihood function $L$ based on Poisson statistics is
\begin{eqnarray}
\ln L = \sum_i \left[ \, C_{\rm obs, \it i} \ln ( C_{\exp, i} ) - C_{\exp, i} \, \right] \ , 
\end{eqnarray}
where the sum is for all cells. 
The expected count $C_{\exp, i}$ is calculated by summing up the expected counts over all pixels in the cell:
\begin{eqnarray}
C_{\exp, i} = \sum_k F_{m, \rm tot}(\Omega_k) \, A_k \, \Delta\Omega_k \, \Delta E \ ,
\end{eqnarray}
where $k$ represents each pixel, $F_{m, \rm tot}$ is the total model diffuse photon flux per unit photon energy and solid angle to the angular direction $\Omega_k$ of the $k$-th pixel, $A_k$ and $\Delta \Omega_k$ are the LAT exposure (in units of $\rm cm^2 \, s$) and the solid angle of the $k$-th pixel, and $\Delta E$ is the width of the photon energy bin under consideration. Though the expected counts should be convolved with the point spread function (PSF), this process is omitted, since the cell size is sufficiently larger than the LAT angular resolution in the photon energy range considered here.

The model diffuse flux $F_{m, \rm tot}$ is a linear combination of various map templates:
\begin{eqnarray}
F_{m, \rm tot}(\Omega_k) = \sum_l f_l \, F_l(\Omega_k) \ , 
\end{eqnarray}
where $F_l(\Omega_k)$ is the template map of the $l$-th model component.
Then the best-fit values and statistical errors of the fitting parameters $f_l$ are obtained by maximizing the likelihood function using the Markov chain Monte Carlo (MCMC) method.

\subsection{Model map templates}

Following the analyses on the Fermi bubbles and the GC GeV excess by the {\it Fermi} LAT team \cite{Ackermann2014-FB,Ackermann2017-GCE}, the following components are considered to fit the gamma-ray maps in the ROI: (1) point sources, (2) diffuse interstellar emission by cosmic-ray interactions, (3) isotropic radiation, (4) Loop I, (5) the Fermi bubbles, and (6) a halo-like emission. The modeling of these components will be explained in order below. However, since the Fermi bubble template will be created during the fitting process in this study, it will be described in section \ref{sec:FB}.

The template of point sources in the LAT 14-year Source Catalog\footnote{\url{https://fermi.gsfc.nasa.gov/ssc/data/access/lat/14yr_catalog/}} (4FGL-DR4, \\ \verb|gll_psc_v35.fit|) \cite{Abdollahi2020,Abdollahi2022} is calculated using the parametrized spectra from the catalog and the energy-dependent LAT point spread function (PSF). 
4FGL-DR4 also provides a catalog of extended sources, in addition to the point sources.
Since the extended sources are not many, circular regions with a radius twice the semi-major axis of the catalog are masked and excluded from the analysis.

Gamma rays produced by cosmic rays propagating in the MW are the most significant component of the Galactic diffuse gamma-ray emission.
Here, we utilize the GALPROP code\footnote{\url{https://galprop.stanford.edu/webrun/}} (ver. 54) \cite{Strong2007,Vladimirov2011,Porter2017}, which numerically solves the generation and propagation of cosmic rays and calculates the gamma-ray intensity resulting from interactions with interstellar gas and radiation fields. 
Among the various GALPROP parameter sets used in the LAT team's study of the Galactic diffuse emission \cite{Ackermann2012-CR},  $^S$L$^Z$6$^R$30$^T$150$^C$2 is adopted as the baseline model in the present work. 
This parameter set is used for the inverse Compton scattering (ICS) component in the Galactic interstellar emission model (GIEM, \verb|gll_iem_v07.fits|)\footnote{\url{https://fermi.gsfc.nasa.gov/ssc/data/access/lat/BackgroundModels.html}} \cite{Acero2016-GIEM,Fermi-LAT2019-GIEM}, which is recommended by the LAT team for normal point source analysis.
The GALPROP model radiation is divided into two components, gas and ICS, which are fitted independently.
The gas component is the sum of pion-decay gamma rays produced by cosmic-ray hadrons, and bremsstrahlung from cosmic-ray electrons and positrons.
The flux ratio between pion decay and bremsstrahlung is fixed to the original model value, making the gas component a single template.
The GALPROP model emission can be further divided into different components, and the fitting results with a larger number of sub-components or different model parameters will be discussed in section \ref{sec:GALPROP}.

Loop I is a giant structure spanning more than 100$^\circ$ in the radio sky \cite{Large1962}, and although it is unclear whether it is nearby supernova remnants or an outflow from the GC \cite{Wolleben2007,Kataoka2013}, it has also been detected by LAT \cite{Casandjian2009}.
We adopt the geometric model of Loop I, which assumes a uniform emissivity within two shells with radii of 50--100 pc, based on the nearby hypothesis. 
The parameters of the geometric model are taken from the LAT team papers \cite{Ackermann2014-FB, Ackermann2017-GCE}, which differ slightly from the values in Wollben's original paper \cite{Wolleben2007}.
The normalization factors for each of the two shells are treated as two independent fit parameters ($f_l$).

To calculate the halo component model map, we first assume that the density profile of the MW dark matter halo is NFW, and use the parameters from the {\it Via Lactea II} simulation \cite{Kuhlen2008,Pieri2011}. 
Therefore, the density profile within the virial radius $r_{\rm vir}$ = 402 kpc is $\rho(r) = \rho_s \, x^{-1} (1+x)^{-2}$, where $x = r/r_s$, $r_s = $ 21 kpc, and $\rho_s = 8.1 \times 10^6 M_\odot \rm \, kpc^{-3}$. 
The distance from the GC to the solar system is set at 8 kpc, and hence $\rho_\odot = 0.42 \ \rm GeV \, cm^{-3}$.
Annihilation gamma-ray diffuse flux is proportional to the line-of-sight integration $\int \epsilon_\gamma \, dl$ of the gamma-ray emissivity $\epsilon_\gamma$ per unit volume. 
If dark matter is distributed smoothly, the emissivity obeys $\epsilon_\gamma \propto \rho^2$, and the map template in this case is denoted as NFW-$\rho^2$.
According to numerical simulations, fine substructures and subhalos exist within halos composed of cold dark matter, and subhalos with high internal densities may dominate annihilation gamma-ray luminosity.
If the number density of subhalos is proportional to the smooth density profile, the gamma-ray emissivity would be $\epsilon_\gamma \propto \rho^1$. 
This emissivity also applies when dark matter particles decay rather than annihilate. 
Therefore, the map template of $\epsilon_\gamma \propto \rho^1$ (NFW-$\rho^1$) will also be tested.
Subhalos may be destroyed by strong tidal forces near the center of the parent halo \cite{Kuhlen2008}, in which case emissivity in the central region would be suppressed.
While a model for this case is not tested in this work, this possibility should be kept in mind.
Previous studies on the GC GeV excess found that the density profile of $\rho \propto r^{-1.25}$ ($\epsilon_\gamma \propto r^{-2.5}$) is preferred, and hence the template of $\epsilon_\gamma \propto \rho^{2.5}$ with the original NFW (NFW-$\rho^{2.5}$) is adopted as the model of the GC excess, whose emissivity becomes $\epsilon_\gamma \propto r^{-2.5}$ in the central region ($\rho \propto r^{-1}$).

The map morphologies of the point sources and GALPROP have photon energy dependencies determined by the models, while other model maps do not depend on photon energy.
The initial values of $f_l$ in the MCMC chain are the same as the original normalization for the models of point sources and GALPROP. 
The initial $f_l$ of the isotropic component is set to $E^2 \, dN/dE = 10^{-4} \ \rm MeV \, cm^{-2} \, s^{-1} \, sr^{-1}$, which is typical of the reported fluxes of the unresolved extragalactic gamma-ray background radiation \cite{Ackermann2015-EGRB}. For other components, the initial values of the MCMC chain are set to zero.
During the MCMC fitting, the fitting parameters are limited to the range $f_l  > 0$ for known components such as point sources, the GALPROP models, the isotropic component, and Loop I, but $f_l < 0$ is allowed for the halo components, which are explored as unknown components.

\section{Search for a halo-like excess}
\label{sec:results}

\subsection{Construction of the Fermi bubble template}
\label{sec:FB}

Unlike Loop I and the MW halo components, there are no parameterized analytical models for the Fermi bubbles. 
Therefore, the map template for the bubbles used in this study is made here, mostly following the methods adopted in the LAT team's papers \cite{Ackermann2014-FB,Ackermann2017-GCE}.
Two photon energy bins, 1.5 and 4.3 GeV, are selected for this procedure, which are in the typical photon energy range where the Fermi bubbles are prominent.
As the first step, a flat (i.e., isotropic) template with sharp edges within a rough area of the Fermi bubbles is assumed, and it is fit to the data together with the other model components.  
Here, the GC GeV excess is included in the fit as a known component, but other halo components are not considered.
At this stage, the fit is performed in the whole ROI, including the Galactic disk region ($|b| < 10^\circ$), to ensure an accurate fit to the GC excess. 

\begin{figure}[tbp]
\centering 
\includegraphics[width=1.0\textwidth]{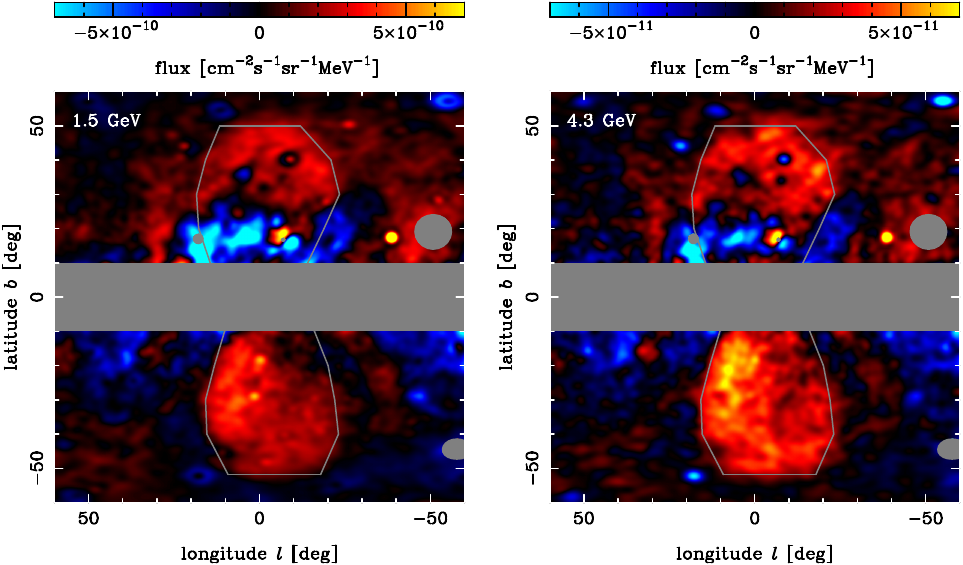}
\caption{\label{fig:map_FB} 
The image of the Fermi bubbles (the flat bubble template plus fit residuals, after Gaussian smoothing with $\sigma = 1^\circ$) in the two photon energy bins (1.5 and 4.3 GeV). 
The boundaries of the final flat template are indicated by gray lines. 
The gray circular regions are excluded from the analysis due to the catalogued extended sources.
The large regions at $(l, b) \sim (-50^\circ, 20^\circ)$ and $(-60^\circ, -45^\circ)$ are Cen A lobes and the SMC, respectively.}
\end{figure}

The best-fit image of the bubble is created from the sum of the flat template and fit residuals; the photon count map of this is converted into physical flux at each pixel using the exposure data, and then smoothed with a Gaussian filter with $\sigma = 1^\circ$.
The energy-independent boundary of the flat template is iteratively improved by looking at these best-fit images. 
The image of the Fermi bubbles obtained in this way in the two energy bins is shown in figure \ref{fig:map_FB}, together with the final boundary of the flat template.
Since the template of the Fermi bubbles will only be used at $|b| \ge 10^\circ$ in subsequent halo component searches, the Galactic disk region ($|b| < 10^\circ$) is not shown.
Residuals of bright point sources are visible in some places, and these are likely due to uncertainties in the catalog spectral models and PSF. 
The fit in this study is performed on a cell-by-cell basis with the cell size of $\Delta l_c = \Delta b_c = 10^\circ$, and PSF errors should have little effect on the results.

\begin{figure}[tbp]
\centering 
\includegraphics[width=.75\textwidth]{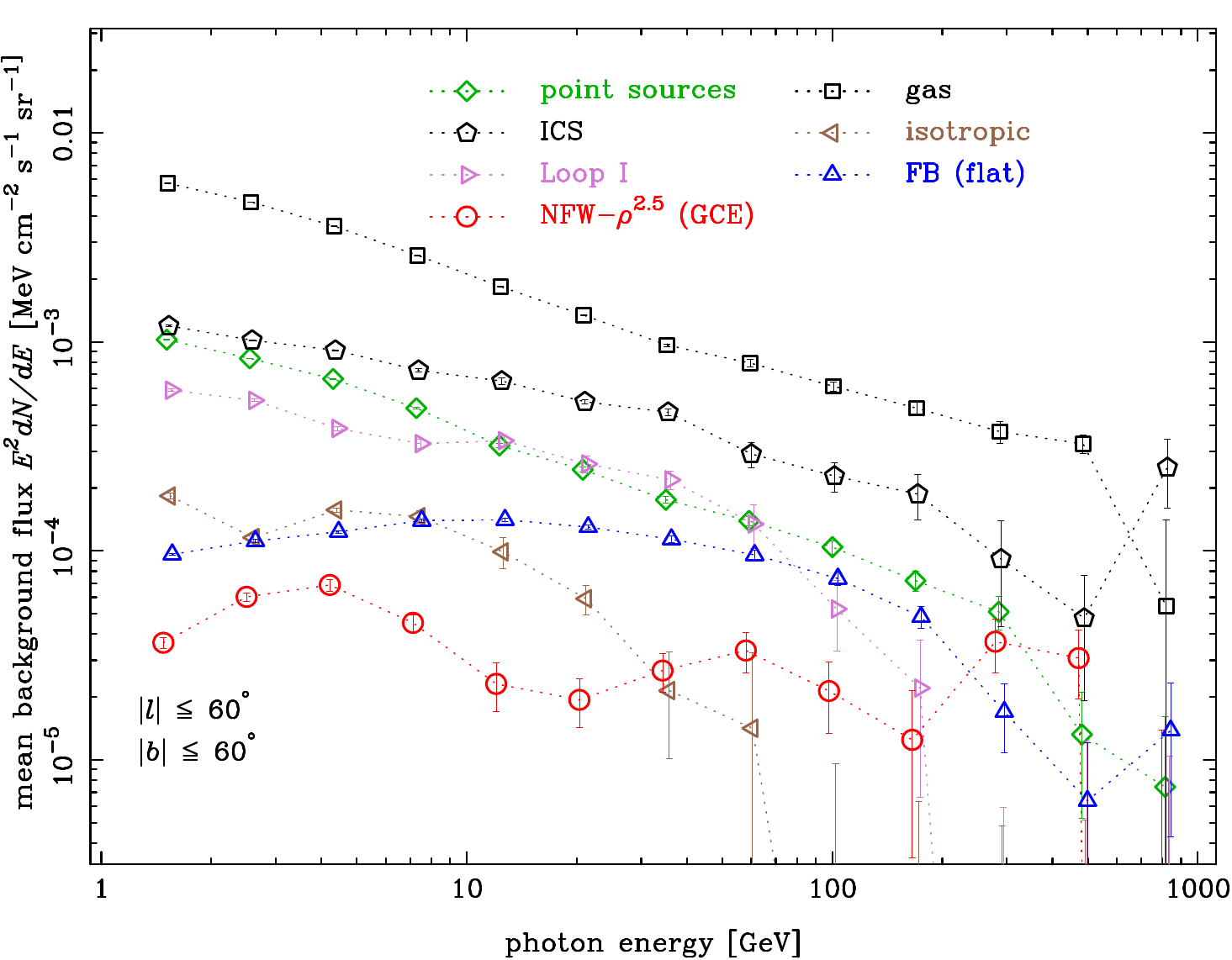}
\caption{\label{fig:specall_FB_b0} 
The best-fit spectra of model map templates, for the fit including the flat template of the Fermi bubbles (FB) and the GC GeV excess (GCE). The error bars are statistical 1$\sigma$ estimated by MCMC.  
The fit is performed including the Galactic disk region ($|b| < 10^\circ$), and the mean background flux within the ROI, including the disk, is shown. 
To prevent symbols and error bars from overlapping, some symbols are slightly shifted to the left or right of the true energy bin center.}
\end{figure}

\begin{figure}[tbp]
\centering 
\includegraphics[width=.75\textwidth]{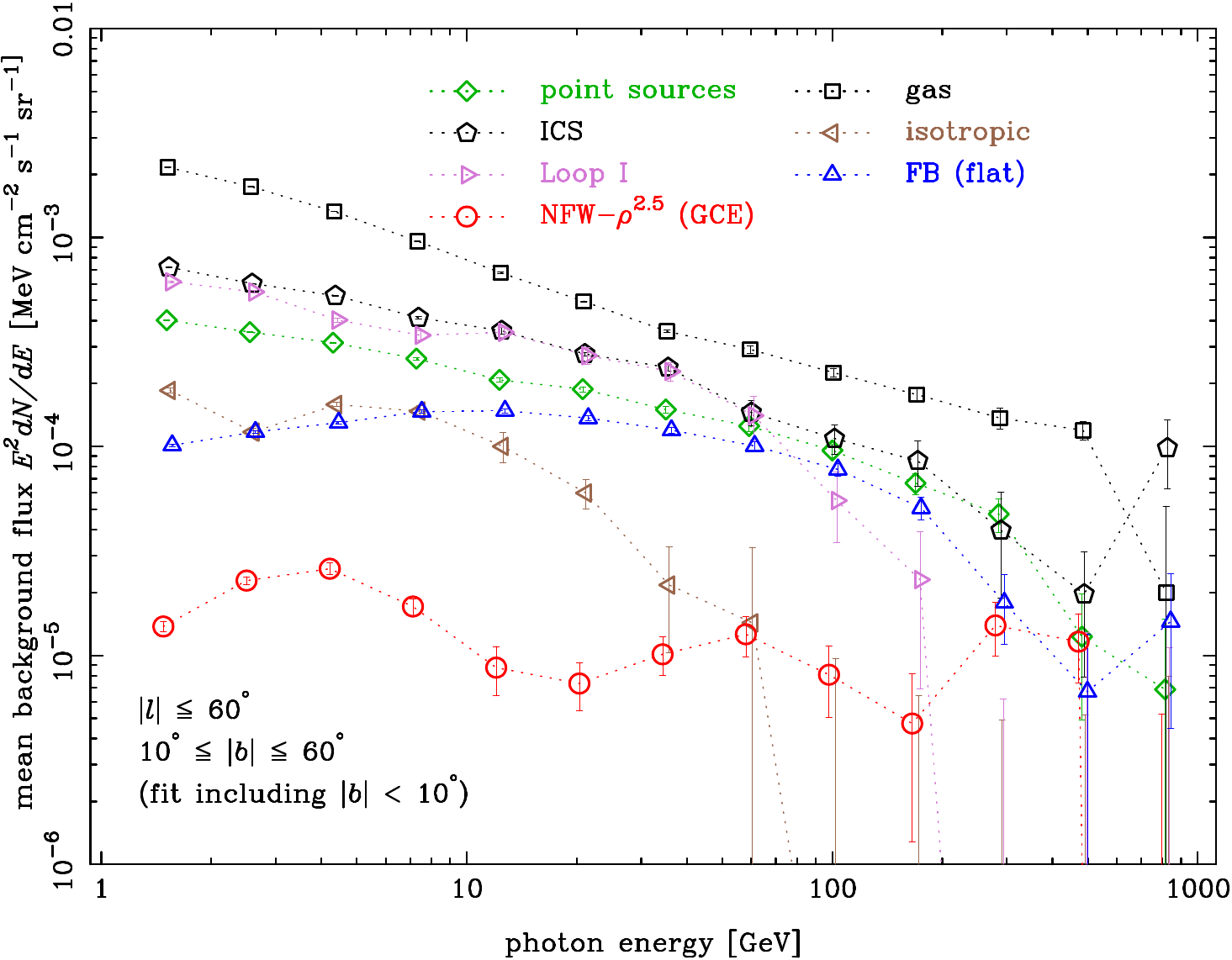}
\caption{\label{fig:specall_FB} 
The same as figure \ref{fig:specall_FB_b0}, but showing the flux excluding the Galactic disk region ($|b| < 10^\circ$), while the fitting is the same as figure \ref{fig:specall_FB_b0} (including the disk). }
\end{figure}

The best-fit spectra of all the model components by the fit with the final flat bubble template are presented in figure \ref{fig:specall_FB_b0}. Here,  the mean background flux within the ROI, including the Galactic disk region, is shown. 
For comparison with later analysis, figure \ref{fig:specall_FB} shows the flux excluding the disk, though the model fitting is the same as figure \ref{fig:specall_FB_b0}, i.e., including the disk. 
The spectra of the model templates are generally consistent with those of similar analyses by the LAT team \cite{Ackermann2017-GCE}, indicating that there are no serious problems or errors in this analysis.
In particular, the peak photon energy and the flux of the GC GeV excess ($2.5 \times 10^{-9} \ \rm cm^{-2} \, s^{-1} \, sr^{-1} \, MeV^{-1}$ at 1$^\circ$ from the GC at 2.5 GeV) are also consistent with those reported in the literature \cite{Murgia2020}.

The morphology of the Fermi bubbles is also consistent with what has been reported in the literature. 
However, in the northern region ($b \sim$ 10--15$^\circ$), negative residuals are clearly visible, especially at the lower energy bin of 1.5 GeV.
This feature corresponds to the regions of particularly strong flux in the gas component map of GALPROP (see figure \ref{fig:map_models} in section \ref{sec:morphology} later), and is likely due to discrepancies between the GALPROP model and the data.
A similar feature is seen in the LAT team's analysis using GALPROP (figure 22 of ref. \cite{Ackermann2014-FB}), and the LAT team used the method of spectral component analysis to produce a bubble template without this feature.
Since the purpose of the present study is not to create an accurate template of the Fermi bubbles, the following approach will be taken to explore an additional halo component.

Among the two energy bins, the 4.3 GeV map is chosen as the source to create the structured bubble template, because the Fermi bubbles are more clearly seen. 
In the map shown in figure \ref{fig:map_FB}, the positive flux regions are dominated by the Fermi bubbles. 
Therefore, the map of positive regions, regardless of whether it is inside or outside the boundaries of the flat template, will be used as the energy-independent template for the bubbles.
On the other hand, to minimize the impact of the discrepancy between GALPROP and the data, the map of the negative flux regions is introduced into the fit as another template.
Since these two probably have different energy spectra, they are included as two independent templates. 
Hereafter, they will be referred to as the positive and negative residual templates, though these strictly include not only the fit residual but also the flat bubble template.

With this treatment, the model and data maps should agree well without additional model components at around 4.3 GeV.
Even if a halo component is present near this energy bin, it will be absorbed by these templates.
However, we are interested in a halo-like emission with a characteristic photon-energy dependence. 
If such a component really exists, we expect to see positive or negative excesses of the model halo template in other energy regions. 
This is the signal we will be searching for.

\subsection{Fitting with a halo-like component}
\label{sec:fit_halo}

From here, the fits will be performed excluding the Galactic disk region ($|b| < 10^\circ$) to search for a halo-like component. 
The flat template of the Fermi bubbles is replaced with the positive and negative residual templates obtained in the previous section.
The fit parameter $f_l$ is limited to $f_l \ge 0$ for the positive residual template (the Fermi bubbles), but not for the negative template, because the inverted negative template (i.e., positive flux) may be favored for a better fit, if the true morphology of the Fermi bubbles is relatively flat and the discrepancy between GALPROP and the data decreases in higher energy bins.

\begin{figure}[tbp]
\centering 
\includegraphics[width=.75\textwidth]{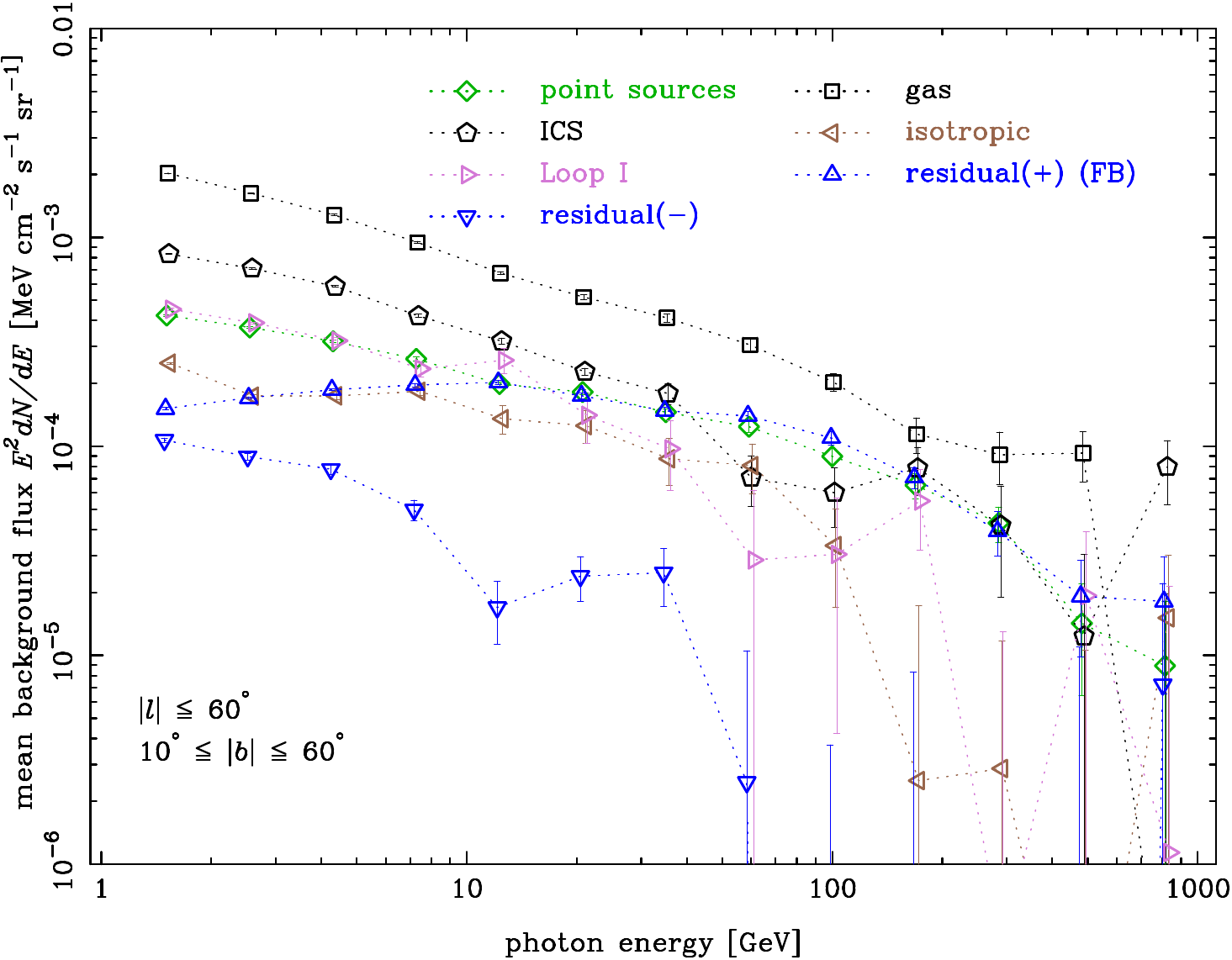}
\caption{\label{fig:specall_noDM} 
The best-fit spectra of the model map templates. 
Compared to figure \ref{fig:specall_FB}, the Galactic disk region ($|b| < 10^\circ$) is removed in the fit, the flat template of the Fermi bubbles is replaced by the two (positive and negative) residual maps, and the GC excess is not included. The sign of the negative template is inverted so that the flux becomes positive in this plot.}
\end{figure}

First, the fit is performed without the GC excess or any other halo component, and the best-fit spectra are shown in figure \ref{fig:specall_noDM}.
The spectrum of the new Fermi bubble template (positive residual map) is almost the same as that of the flat template, and as expected, the negative template spectrum is softer than the Fermi bubbles and closely resembles that of the gas component.
Another notable point is that the ICS component is softer than the fit including the Galactic plane (figure \ref{fig:specall_FB}). 
This suggests that, compared to the GALPROP model, the ICS spectrum softens from the Galactic plane to high Galactic latitudes.

\begin{figure}[tbp]
\centering 
\includegraphics[width=.75\textwidth]{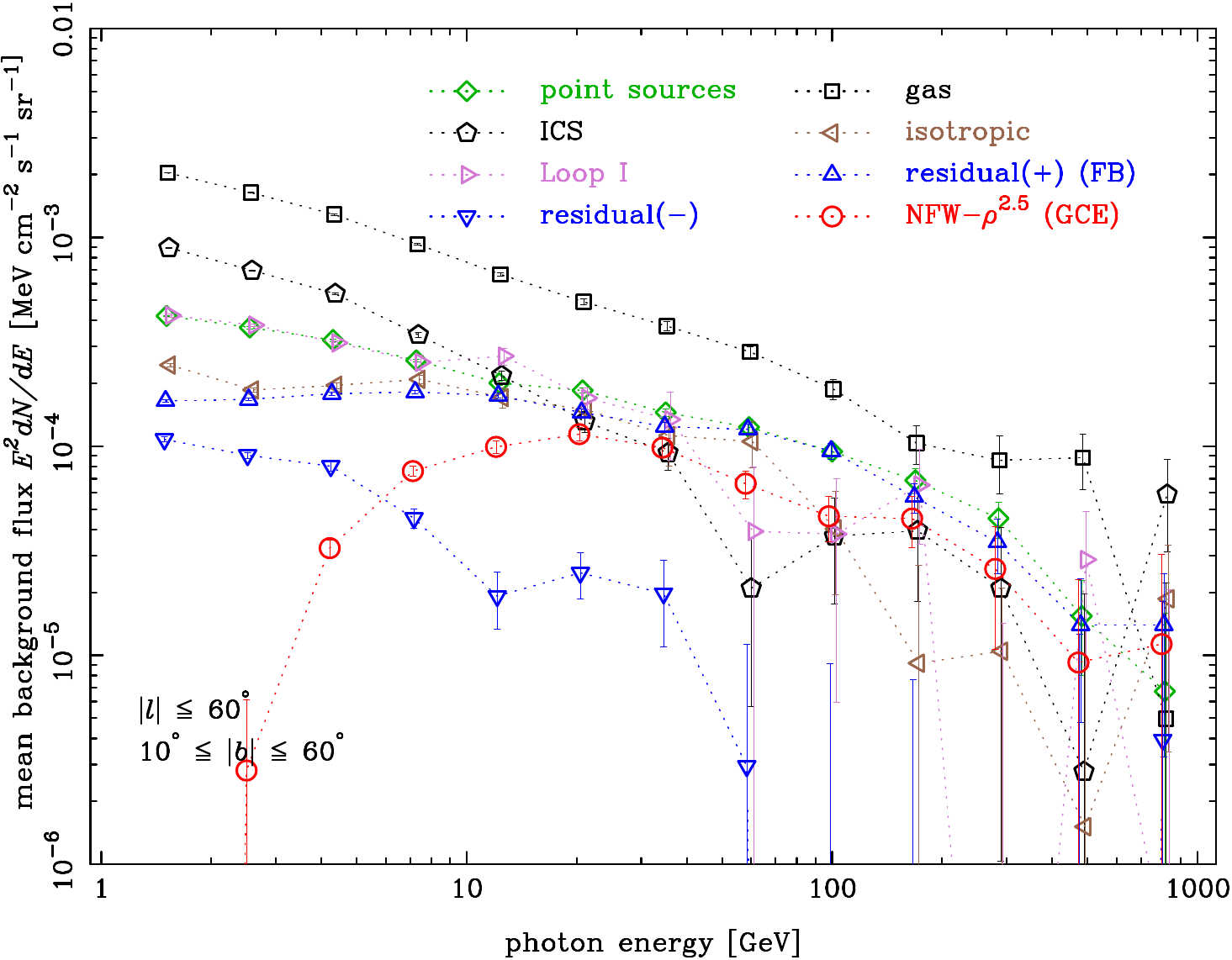}
\caption{\label{fig:specall_GCE} 
The same as figure \ref{fig:specall_noDM}, but including the GC GeV excess (GCE) template (NFW-$\rho^{2.5}$). 
}
\end{figure}

Next, the fit results with the GC GeV excess component (NFW-$\rho^{2.5}$) added are shown in figure \ref{fig:specall_GCE}.
The GC excess flux is the same as the previous fit that includes the Galactic plane (figure \ref{fig:specall_FB}) at 4.3 GeV. 
This is how it should be, because the two residual templates are created at this energy bin. 
However, in higher photon energy regions, the GC excess component rises sharply and peaks at around 20 GeV, where the flux is more than 10 times higher than the previous fit. 
This means that extrapolating this fit to the GC would significantly exceed the observed data.
These results imply that a halo-like component exists around 20 GeV and is prominent in the $|b| > 10^\circ$ region, with a density profile shallower than that assumed for the GC GeV excess.

\begin{figure}[tbp]
\centering 
\includegraphics[width=.75\textwidth]{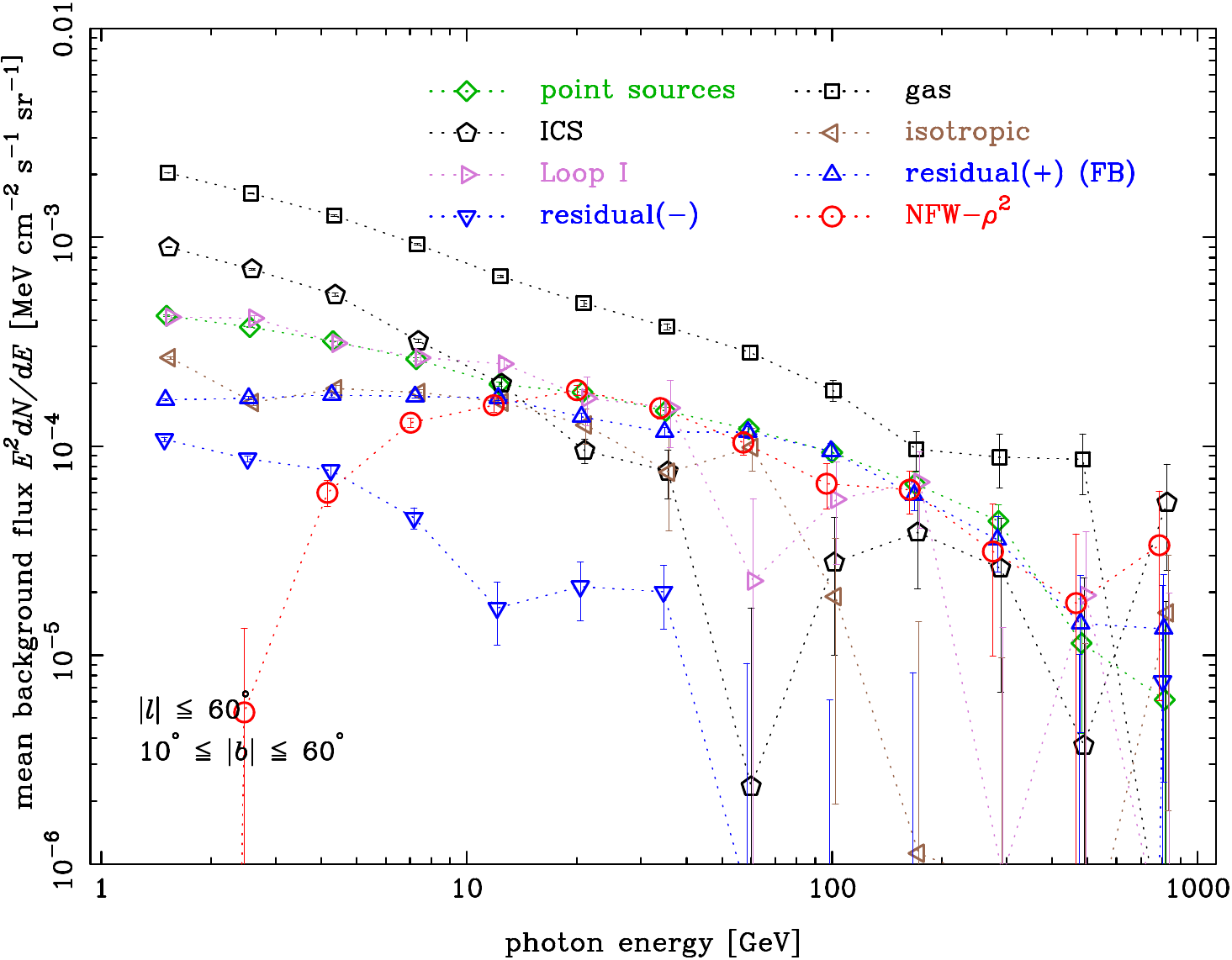}
\caption{\label{fig:specall_NFW2} 
The same as figure \ref{fig:specall_GCE}, but the GC GeV excess template is replaced by the NFW-$\rho^2$ model (NFW and $\epsilon_\gamma \propto \rho^2$). }
\end{figure}

\begin{figure}[tbp]
\centering 
\includegraphics[width=.75\textwidth]{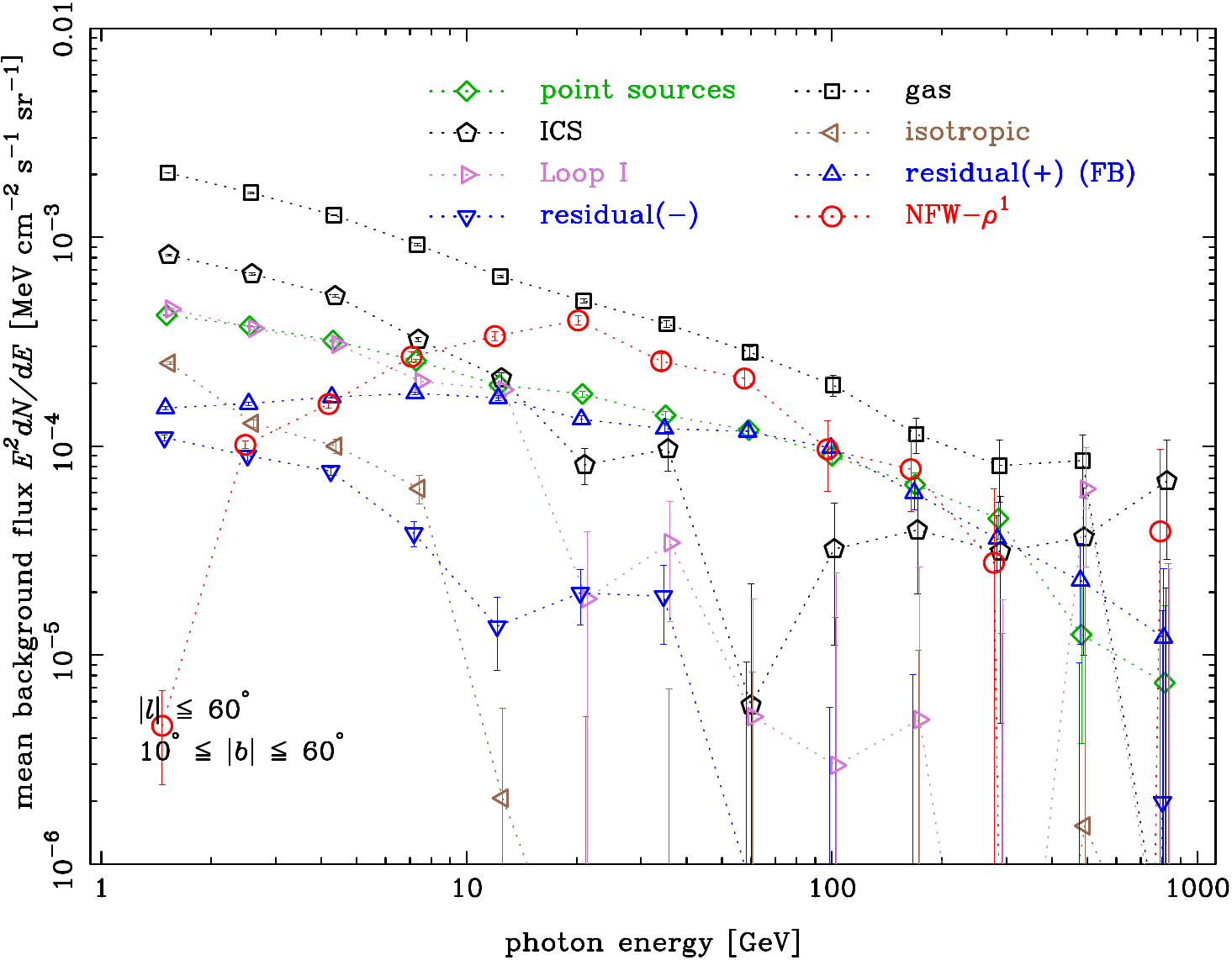}
\caption{\label{fig:specall_NFW1} 
The same as figure \ref{fig:specall_GCE}, but the GC excess template is replaced by the NFW-$\rho^1$ model (NFW and $\epsilon_\gamma \propto \rho^1$).  }
\end{figure}

\begin{figure}[tbp]
\centering 
\includegraphics[width=1.0\textwidth]{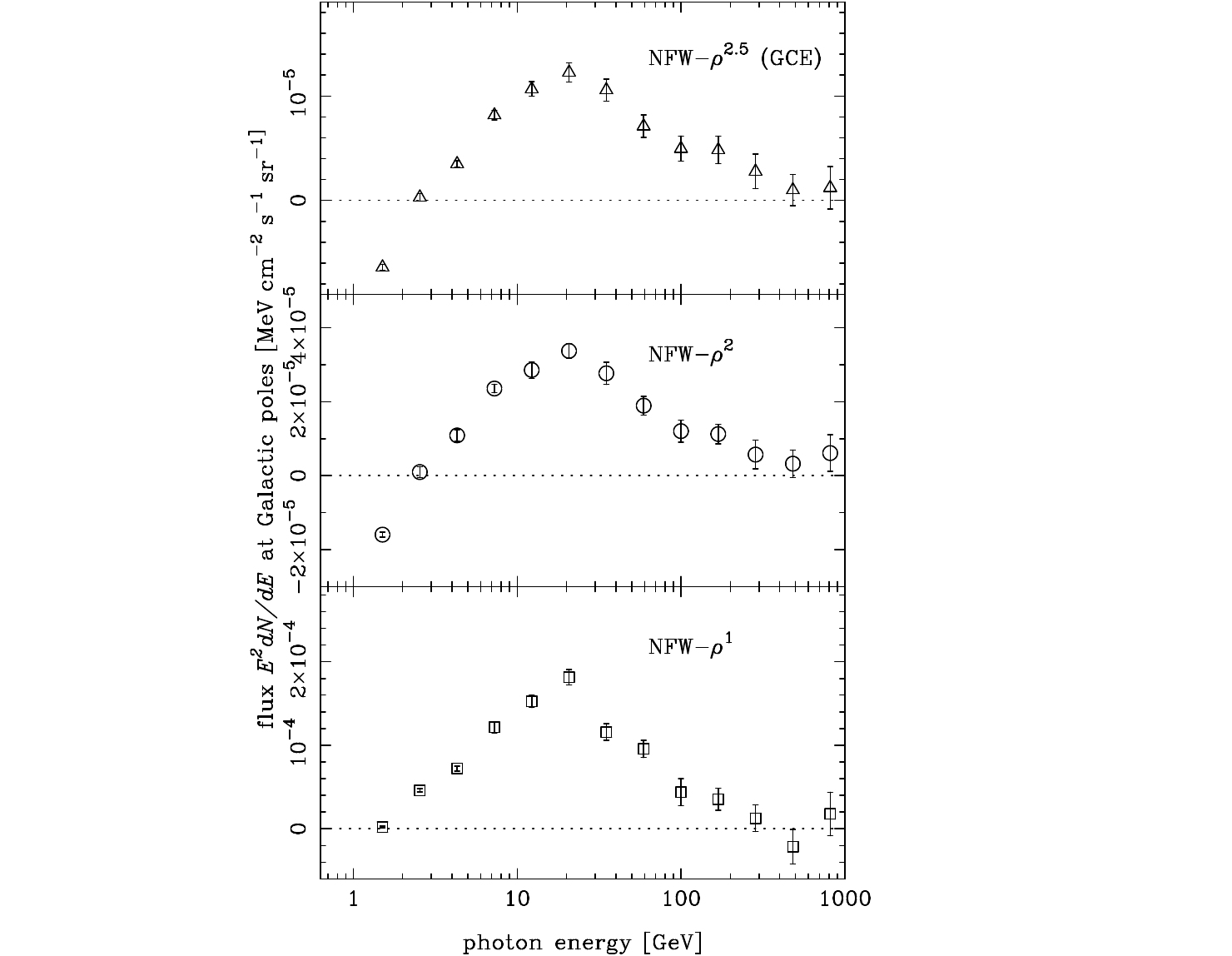}
\caption{\label{fig:spec_NFWX} 
The spectra of the halo model components (NFW-$\rho^{2.5}$, -$\rho^{2}$, and -$\rho^{1}$) in figures \ref{fig:specall_GCE}, \ref{fig:specall_NFW2}, and \ref{fig:specall_NFW1} are shown in linear plots.
The model diffuse fluxes at the Galactic poles ($b = \pm 90^\circ$) are shown. 
}
\end{figure}

Therefore, to test a flatter density profile, we perform fits using the NFW-$\rho^2$ and NFW-$\rho^1$ halo models instead of the GC excess, and the results are shown in figures \ref{fig:specall_NFW2} and \ref{fig:specall_NFW1}, respectively. 
Only one of the three halo models is used in a fit to avoid degeneracy between the similar halo maps.
Since the best-fit halo flux becomes negative at the lowest photon energy bin, figure \ref{fig:spec_NFWX} shows the spectra of the three halo models on a linear plot.
The flatter the density profile, the higher the mean background flux in the ROI, but in all cases, the intensity of the halo components rises sharply with photon energy between 1 GeV and 20 GeV, reaching a statistically significant (13--19$\sigma$) peak at the 21 GeV bin.
At higher energy bins, the halo flux decreases rapidly, becoming almost zero at 200 GeV or higher.

The halo components are not completely zero at the 4.3 GeV bin, where the positive and negative templates are created from the residual map, because the GC excess component was included in the fit that yielded the residual map. 
The halo component flux decreases toward lower photon energies, becoming negative at 1.5 GeV. 
This may be attributed to the difference between the GALPROP gas component and the data, which becomes more pronounced at lower energies.
Another possibility is that part of the halo component at 4.3 GeV has been absorbed in the residual map templates.
If the halo component weakens toward lower energy below 4.3 GeV, the fit of the halo component could be negative to cancel out the part absorbed in the residual map templates. 

\begin{figure}[tbp]
\centering 
\includegraphics[width=.73\textwidth]{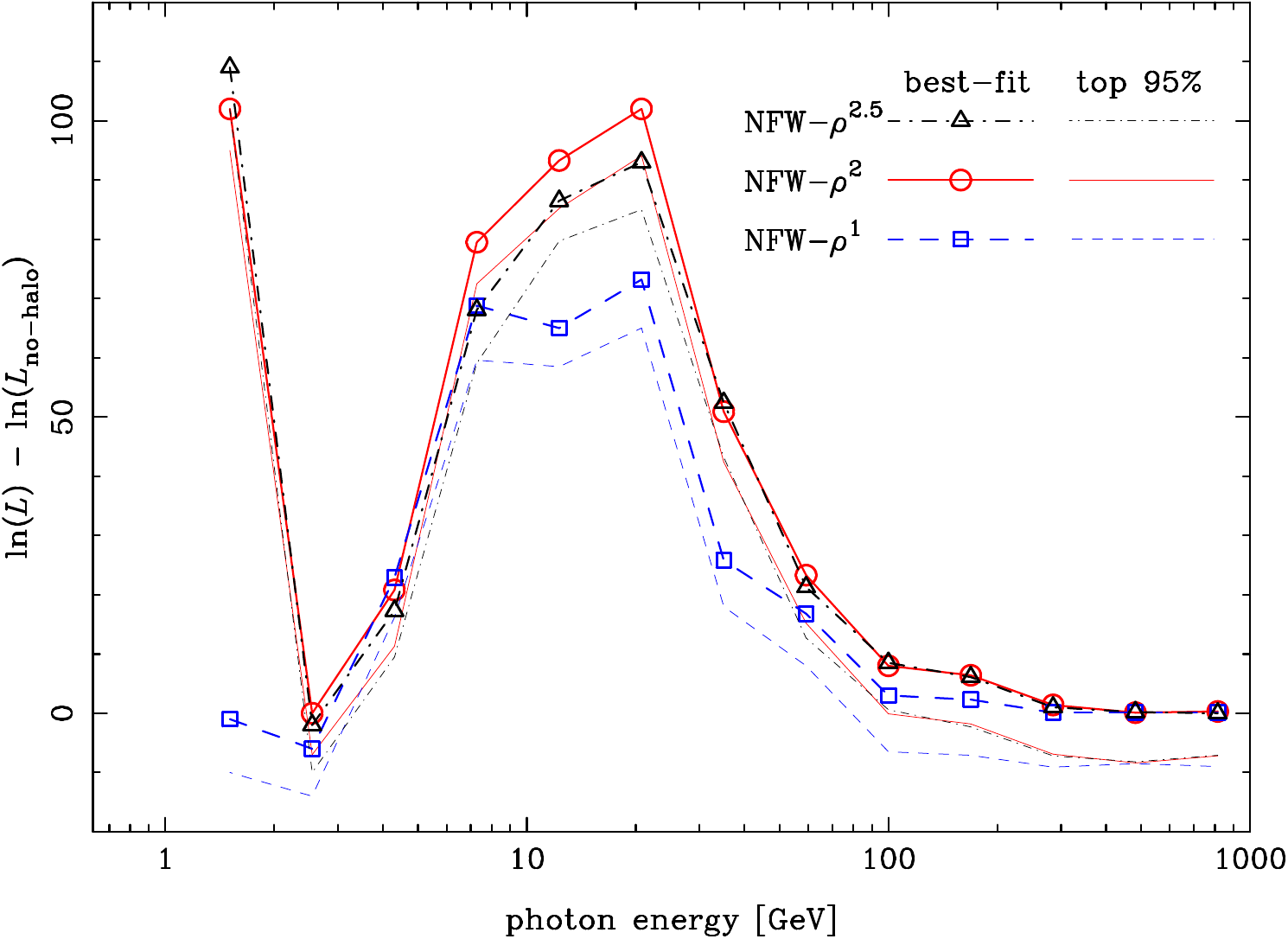}
\caption{\label{fig:like} The log-likelihood differences of the fits by the three halo models, from the best-fit likelihood of the no-halo fit. The thick lines with symbols represent the best fit (the maximum likelihood), while the thin lines represent the boundary that includes the top 95\% in the MCMC chain.
}
\end{figure}

Figure \ref{fig:like} illustrates the likelihood (relative to the no-halo fit) of the fits with three different halo models, to compare the goodness of fit.
In the 2.5 GeV bin, the halo components are fitted to be nearly zero (figure \ref{fig:spec_NFWX}), and a deep dip is seen in the figure.
Around the peak photon energy of the halo excess (10--20 GeV), the NFW-$\rho^2$ model fits better than the steepest NFW-$\rho^{2.5}$ at $\sim$95\% confidence level (CL).
On the other hand, the shallowest NFW-$\rho^1$ model fits even worse than the NFW-$\rho^{2.5}$ (at a significance much higher than 95\% CL), suggesting that a profile close to NFW-$\rho^2$ best fits the data in the region of $|b| \ge 10^\circ$.

\subsection{Morphology and radial profile of the halo-like excess}
\label{sec:morphology}

\begin{figure}[tbp]
\centering 
\includegraphics[width=.73\textwidth]{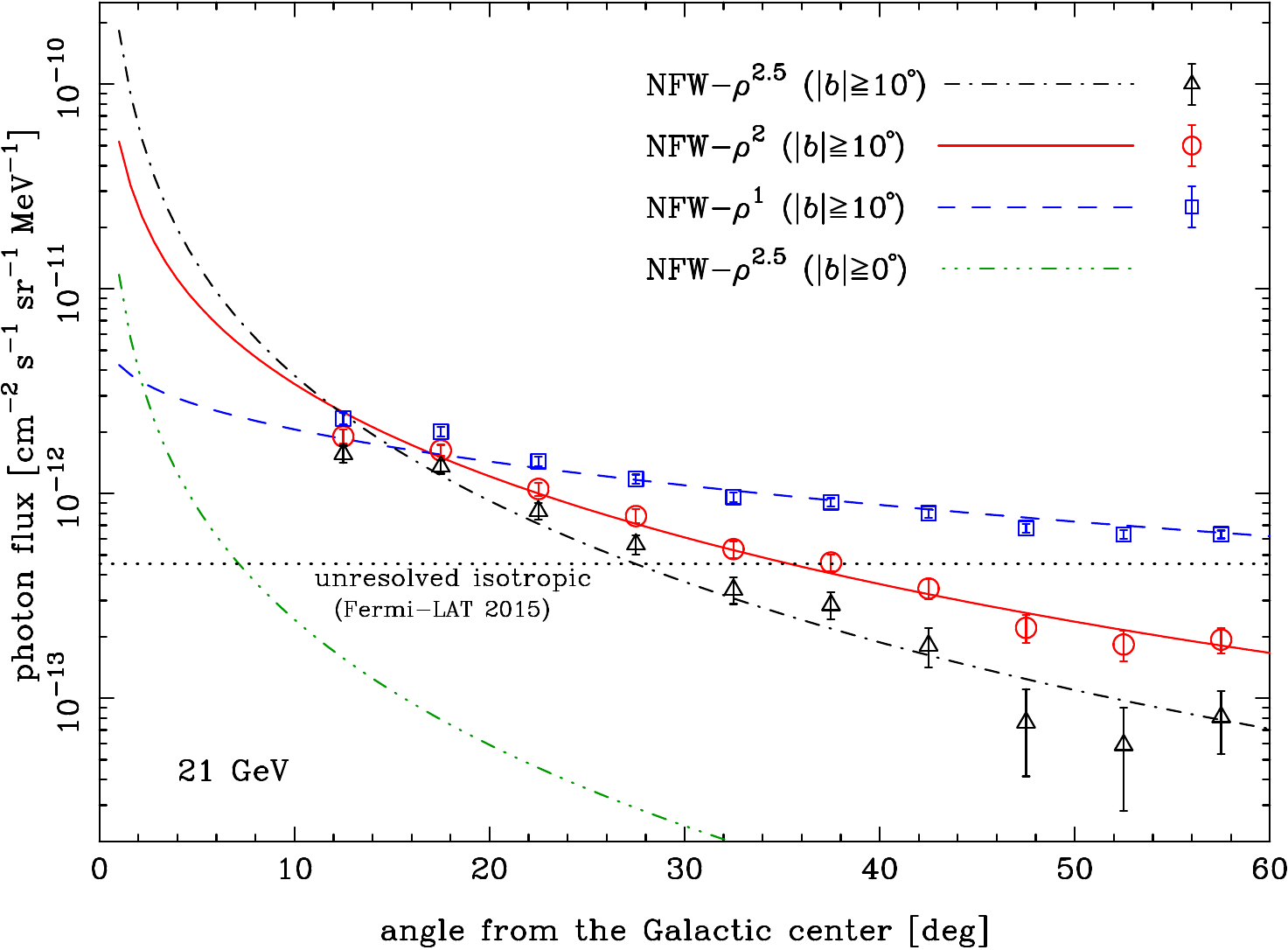}
\caption{\label{fig:fl_angle} 
The radial angular profiles are shown for the best-fit halo models (curves) in the 21 GeV bin. 
The curve for NFW-$\rho^{2.5} \ (|b| \ge 0^\circ)$ is the fit for the GC GeV excess in the ROI including the Galactic plane ($|b| < 10^\circ$), while the other three curves and data points are the fits excluding the Galactic plane.
The data points are the sum of the best-fit model and the fit residuals, and the 1$\sigma$ statistical errors are Poisson statistics of the total observed photon counts in an angular bin. 
The unresolved isotropic background flux reported by the {\it Fermi} LAT team \cite{Ackermann2015-EGRB} is shown as the horizontal dotted line. 
}
\end{figure}

Figure \ref{fig:fl_angle} shows the radial angular profiles of three halo model fits at the energy bin of 21 GeV, where the halo component is the most prominent. 
The diffuse flux of the best-fit halo model is shown as a curve, and the data points are the sum of the best-fit model and the fit residuals, in a ring region centered on the GC within the ROI ($|l| \le 60^\circ$, $10^\circ \le |b| \le 60^\circ$).  It is not straightforward to determine from this figure which of the three halo models fits best, but as seen in the previous section \ref{sec:fit_halo} and figure \ref{fig:like}, the likelihood values indicate that the NFW-$\rho^2$ model is preferred with a statistical significance of 2$\sigma$ or greater. 
The density profile of the NFW-$\rho^{2.5}$ model is steeper than the data, while the NFW-$\rho^1$ model is shallower.

The data points differ significantly among the three models, especially in the tail region, suggesting the presence of degeneracy with other model components. 
In particular, the tail of the halo models is considered prone to degeneracy with the isotropic component.
Indeed, in the NFW-$\rho^1$ fit, the isotropic component decreases significantly near 20 GeV compared to the NFW-$\rho^2$ fit, becoming consistent with zero (figures \ref{fig:specall_NFW2} and \ref{fig:specall_NFW1}). 
In the NFW-$\rho^2$ fit, the flux of the isotropic component near 20 GeV (figure \ref{fig:specall_NFW2}) agrees well with the unresolved isotropic background flux reported by the Fermi-LAT team \cite{Ackermann2015-EGRB}, $E^2 \, dN/dE \sim 2 \times 10^{-4} \ \rm MeV \, cm^{-2} \, s^{-1} \, sr^{-1}$ (shown in figure \ref{fig:fl_angle}).
This implies that the NFW-$\rho^1$ fit leaves no room for the isotropic background that should inherently exist, providing further evidence that the NFW-$\rho^2$ model fits the data better, in addition to the likelihood values.  

For comparison, the best fit flux of the GC GeV excess (the NFW-$\rho^{2.5}$ model in the fit including the Galactic plane, figure \ref{fig:specall_FB_b0}) is also shown in the figure as the curve of ``NFW-$\rho^{2.5}$ ($|b| \ge 0^\circ$)''.
The halo component must have a shallow profile that does not exceed this curve when extrapolated to the GC.
Though the NFW-$\rho^2$ model best fits at $|b| \ge 10^\circ$, it is several times higher than the GC excess at around 1$^\circ$ from the GC.
This suggests that the angular profile that best fits the data across the entire ROI is slightly flatter than NFW-$\rho^2$, or that the angular profile becomes gradually flatter than NFW-$\rho^2$ toward the GC.

\begin{figure}[tbp]
\centering 
\includegraphics[width=1.0\textwidth]{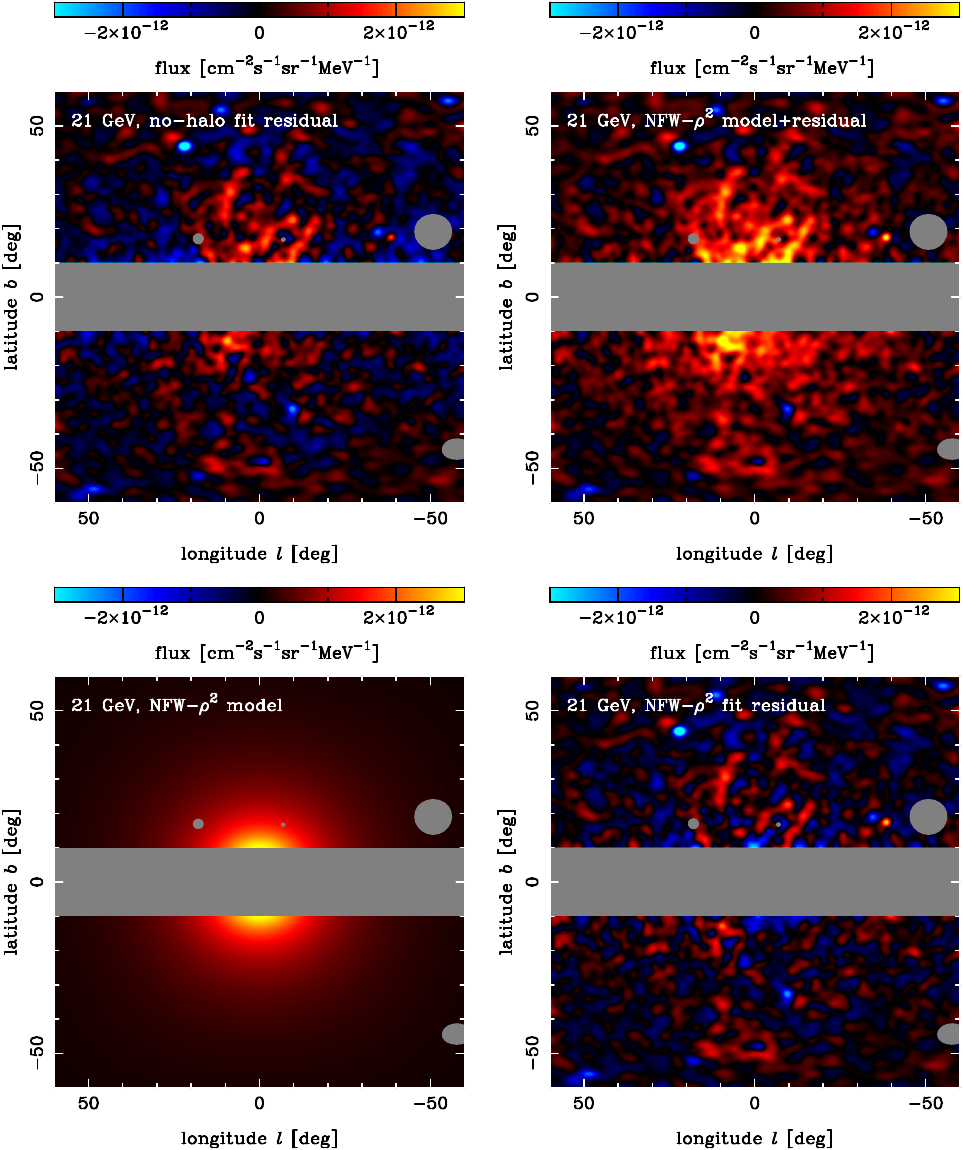}
\caption{\label{fig:map_E12} 
Map images of the fit with the  NFW-$\rho^2$ model at the 21 GeV bin. 
Top left: the residual map of the fit without any halo component.
Top right: the halo model plus residual (i.e., the observed map with all model maps except for the halo subtracted). 
Bottom left: the halo model map.
Bottom right: the residual of the fit with the halo model. 
The maps are smoothed with a Gaussian of $\sigma = 1^\circ$.
See figure \ref{fig:map_FB} caption about the grey circular regions.
}
\vspace*{1cm}
\end{figure}

\begin{figure}[tbp]
\centering 
\includegraphics[width=1.0\textwidth]{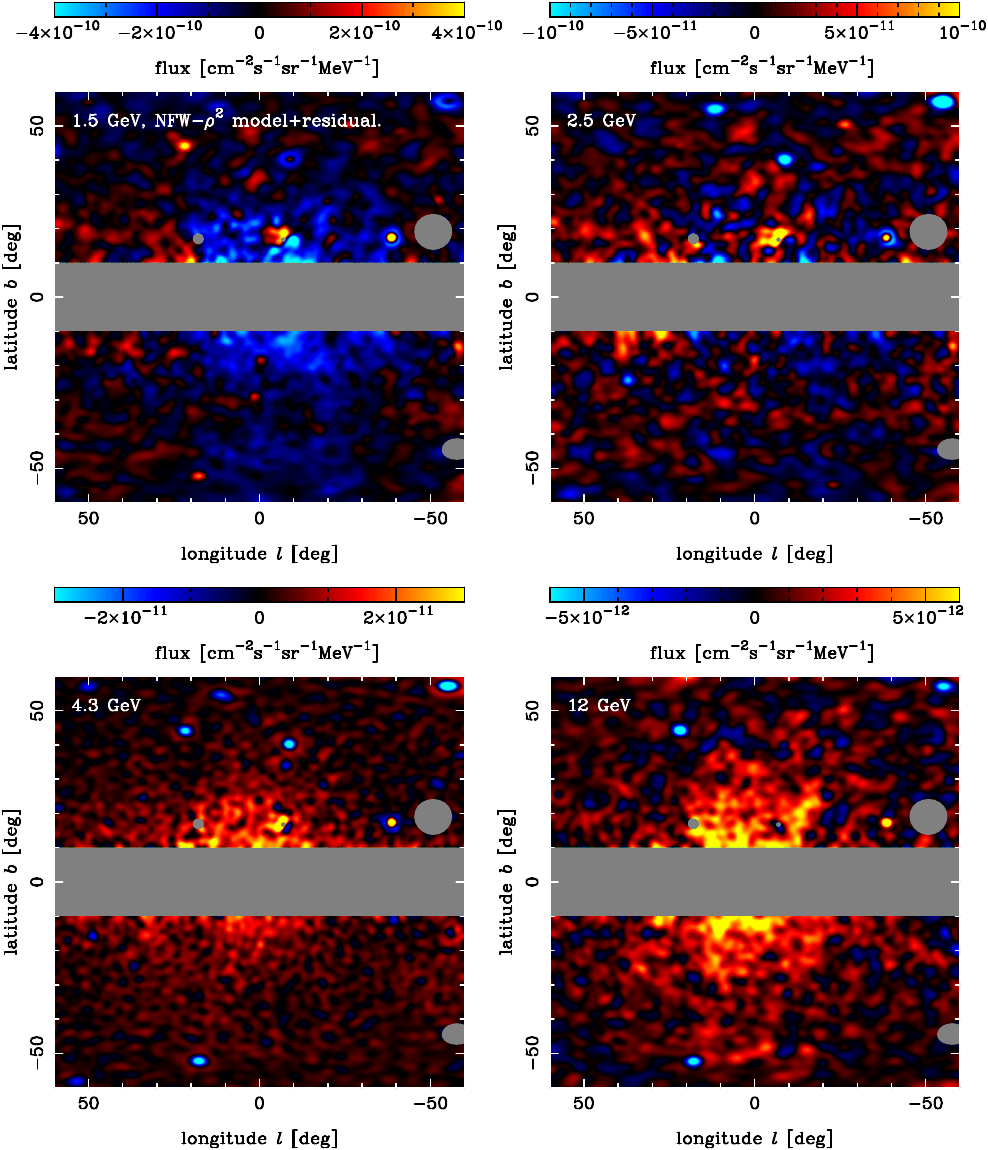}
\caption{\label{fig:map_lowE} 
Map images of the halo model plus residual by the fit with the  NFW-$\rho^2$ model at four photon energy bins lower than 21 GeV. 
The maps are smoothed with a Gaussian of $\sigma = 1^\circ$.
}
\vspace*{3cm}
\end{figure}

\begin{figure}[tbp]
\centering 
\includegraphics[width=1.0\textwidth]{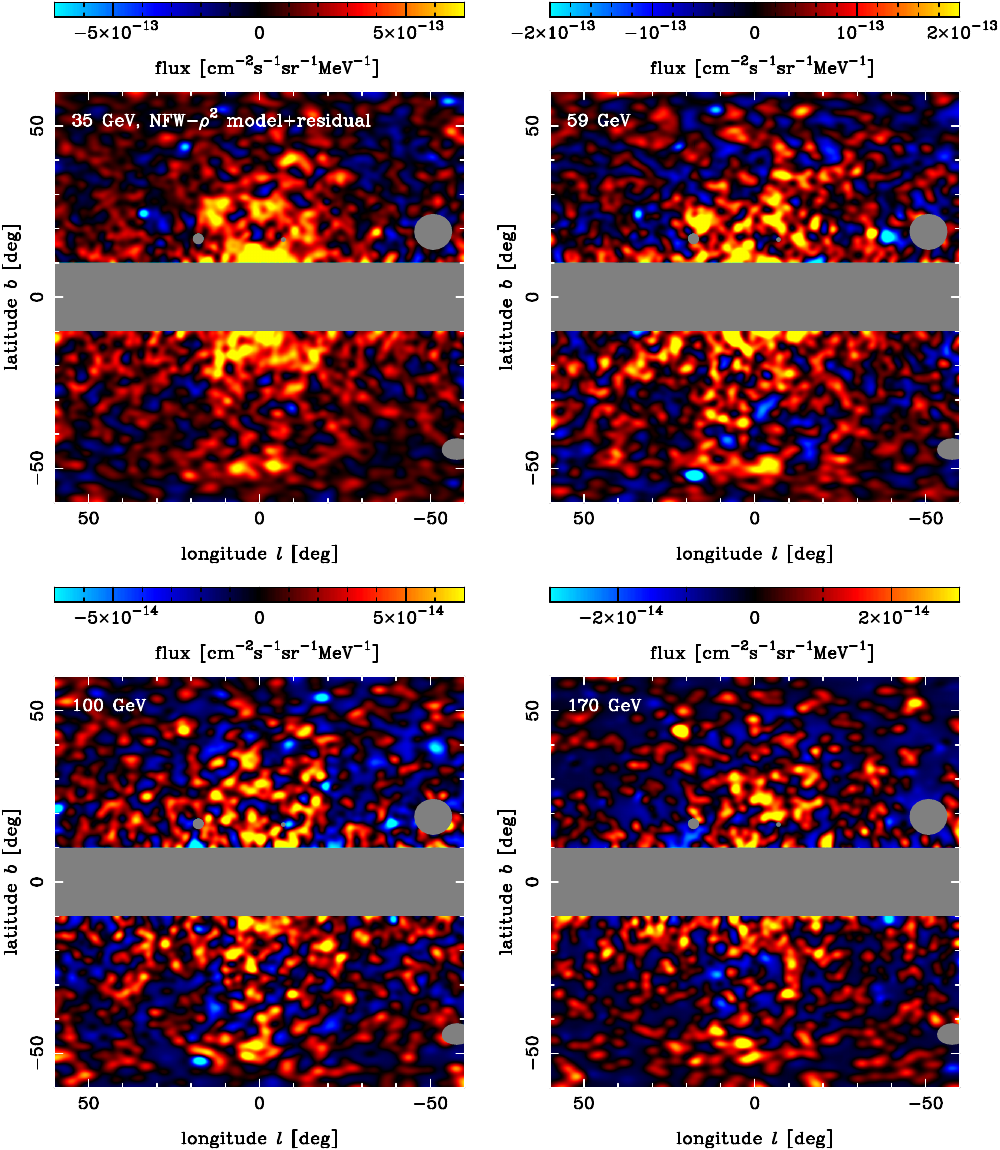}
\caption{\label{fig:map_highE} 
The same as figure \ref{fig:map_lowE}, but for four photon energy bins higher than 21 GeV. 
}
\vspace*{3cm}
\end{figure}

To verify how well the halo model matches the data in terms of morphology, map images are shown in figure \ref{fig:map_E12}, for the NFW-$\rho^2$ fit at the 21 GeV bin.
When the fit is made without a halo model component, the fit tries to minimize the halo component, and hence, the fit would be biased if the halo component physically exists. 
However, a halo-like residual can be seen even in such a fit (top left), giving a lower bound for the halo-like signal. 
In the fit including NFW-$\rho^2$ as the halo model, the map of the model plus residual (i.e., the observed map with all model maps except for the halo subtracted) shows a clearer halo-like structure (top right), which is spherically symmetric and consistent with the model map (bottom left). 
The fit residual with the halo model (bottom right) is significantly reduced compared to the fit without the halo (top left), indicating that the halo fit is successful overall in the ROI. 

To examine the fits in energy bins other than the peak (21 GeV), figures \ref{fig:map_lowE} and \ref{fig:map_highE} show maps of the NFW-$\rho^2$ model plus residual at bins lower and higher than 21 GeV, respectively. 
These maps show the trend seen in figure \ref{fig:spec_NFWX}, with the negative fit at the lowest 1.5 GeV, nearly zero at 2.5 GeV, the halo component becoming stronger toward 21 GeV, and then rapidly weaker beyond 21 GeV. 
The morphologies of the halo model plus residual are generally spherically symmetric and do not appear to contradict the profile of the assumed model halo.

\begin{figure}[tbp]
\centering 
\includegraphics[width=1.0\textwidth]{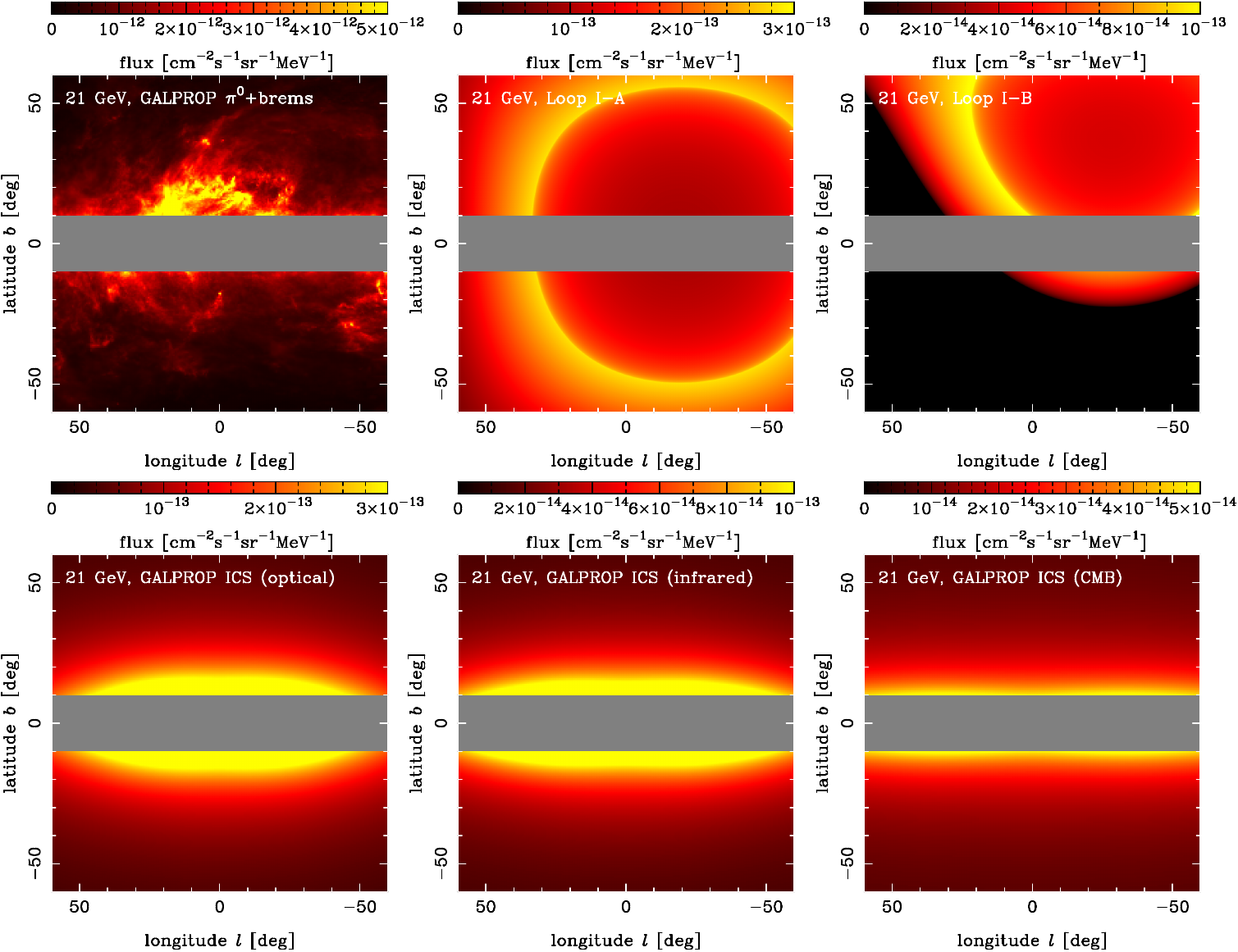}
\caption{\label{fig:map_models} 
Map images of the best-fit model templates (other than the halo) used in the NFW-$\rho^2$ fits, at the 21 GeV bin. 
The top panels show the gas component of GALPROP and the two shells (A and B) of Loop I. 
The bottom panels show the ICS components of GALPROP, divided into the three sub-components of the target photon fields: optical, infrared, and CMB. 
}
\end{figure}

Finally, the images of the best-fit model map templates considered in the fits, other than the halo, are shown in figure \ref{fig:map_models} for comparison. 
The morphology of the GALPROP gas component and Loop I differ greatly from the halo model, and it is unlikely that they are degenerate with each other.
The ICS components of GALPROP are shown separately for each of the three target photon fields: optical (direct radiation from stars), infrared (emission from interstellar dust grains), and the cosmic microwave background (CMB) radiation.
These are shown separately here to check morphology, but it should be noted that they are treated as a single component in the fitting of the baseline analysis.
Though the optical and infrared ICS components show a morphology most similar to that of the halo, they are considerably flattened along the Galactic plane compared with the halo. 
The possible degeneracy between the ICS and halo components will be discussed in section \ref{sec:origins}.

\subsection{Examination of systematic uncertainties}
\label{sec:systematics}

Here, systematic uncertainties concerning the evidence for the halo-like excess are examined on the following points: (1) fit analysis divided into four quadrants, (2) uncertainties in the Fermi bubble template, (3) treatment of point sources, (4) model parameters and number of free components in GALPROP, and (5) halo-like excess search using the {\it Fermi} LAT standard background model (GIEM) as the reference.

\subsubsection{Analysis divided into four quadrants of the Galactic coordinates}

To test the isotropy of the halo excess around the GC, the NFW-$\rho^2$ fits are performed by dividing into four quadrants about the Galactic coordinates, and the halo excess spectra are shown in figure \ref{fig:spec_syst} (top panel).
Furthermore, the fitting results are also illustrated when dividing the four quadrants shifted by 45$^\circ$ into two quadrants in the galactic plane direction ($|l| > |b|$) and two quadrants in the polar direction ($|l| < |b|$).
Although statistical fluctuations increase significantly above 30 GeV due to the decrease in photon number, a consistent trend is observed in all quadrants below 20 GeV.

\subsubsection{The Fermi bubble template}

\begin{figure}[tbp]
\hspace{-2.3cm}
\includegraphics[width=1.4\textwidth]{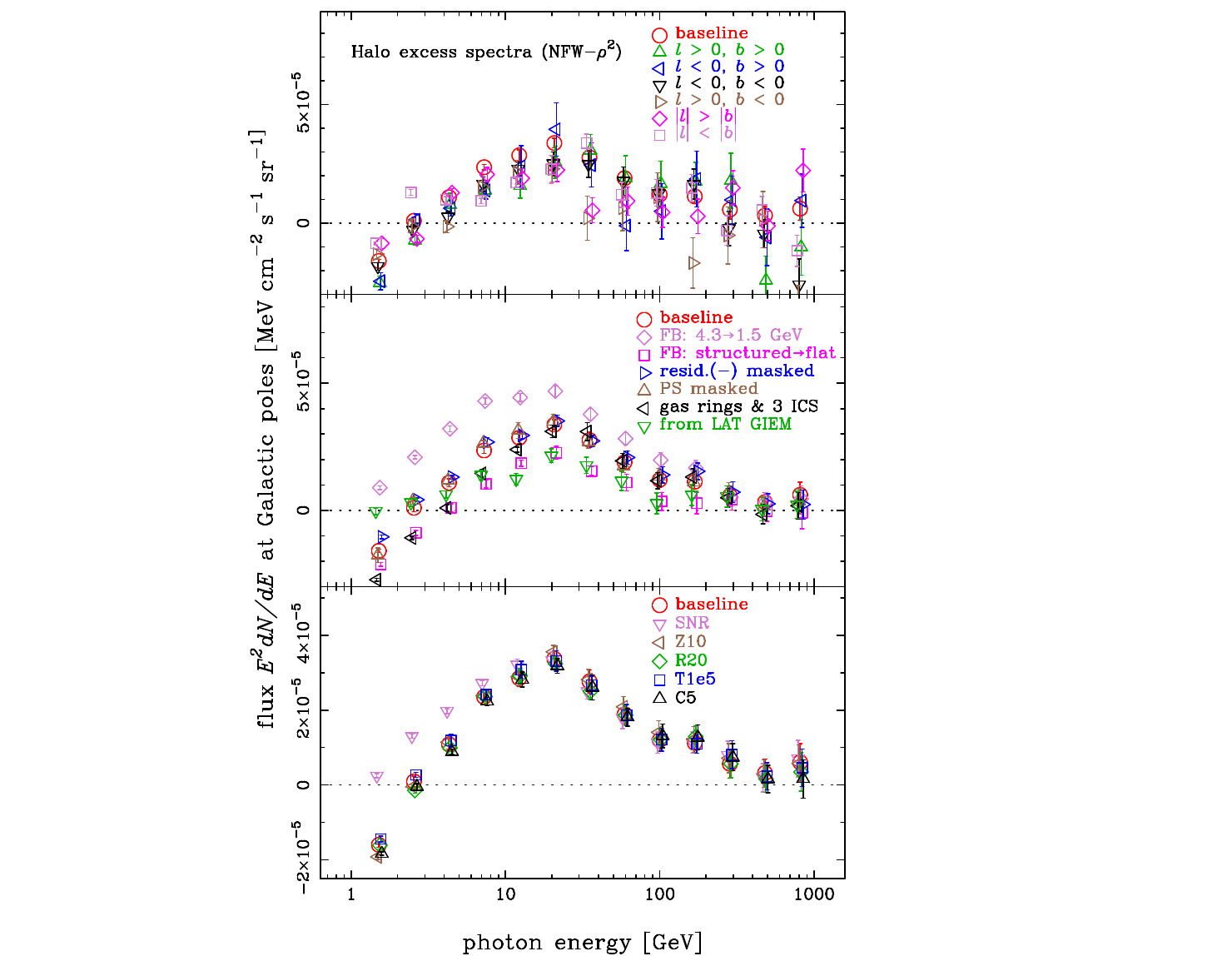}
\caption{\label{fig:spec_syst} 
This figure shows how the halo excess spectrum (the NFW-$\rho^2$ fit) changes from the baseline analysis, depending on model parameters or treatments. 
Top: the spectra by fits using quadrants of the Galactic coordinates are shown.
Middle: ``FB: 4.3$\rightarrow$1.5 GeV'' is when the Fermi bubble template is created from the 1.5 GeV map.
``FB: structured$\rightarrow$flat'' is when the flat template is used for the Fermi bubbles.
``resid.(-) masked'' is the spectrum when the region of negative residuals is masked.
``PS masked'' is when the point source areas are masked instead of model subtraction.
``gas rings \& 3 ICS'' refers to the case where the GALPROP gas component is divided into five Galactocentric rings and the ICS component into three.
``from LAT GIEM'' is when the halo excess is measured with respect to 
the LAT GIEM.
Bottom: results when one of the five GALPROP model parameters is changed.
``SNR'': the SNR distribution for cosmic-ray sources, ``Z10'': $z_h = 10$ kpc, ``R20'': $R_h = 20$ kpc, ``T1e5'': $T_S = 10^5$ K, and ``C5'': the magnitude cut 5 mag for $E(B-V)$.
The errors are statistical 1$\sigma$ estimated by MCMC. 
To prevent symbols and error bars from overlapping, some symbols are slightly shifted to
the left or right of the true energy bin center.
}
\end{figure}

The Fermi bubble template was created from the positive residual map at 4.3 GeV; however, a portion of the halo component in that energy bin may be absorbed by the template, resulting in an underestimate of the halo component in the fit. 
To check the dependence on the choice of energy bin in this process, an alternative fit result is shown in figure \ref{fig:spec_syst} (middle), in which the positive (the Fermi bubbles) and negative residual templates are replaced with those of the 1.5 GeV map shown in figure \ref{fig:map_FB}.
As expected, the energy bin where the halo component becomes zero has shifted to the lower side, and the halo flux at the 20 GeV peak has increased, but there is no change in the basic shape of the spectrum with a peak near 20 GeV.

As shown in figure \ref{fig:map_FB}, due to the strong negative residual around $b \sim$ 10--15$^\circ$, the positive residual template may be biased from the true structure of the Fermi bubbles.
To test the dependence on the Fermi bubble template structure, the fit result obtained by replacing the positive residual template with the simple flat template is shown in figure \ref{fig:spec_syst} (middle).
The halo component has decreased compared to the baseline analysis, likely because the bubble flux in the negative residual region becomes larger compared to the structured bubble template, and the halo component is fitted smaller to compensate for this. 
However, the overall spectral shape, including the rising rate of the halo component from 1 to 20 GeV, remains unchanged. 
The excess at the 21 GeV bin is still statistically significant (9.1$\sigma$).

As a further verification of systematic uncertainties related to the negative residual region, the fit is performed after this region is masked. 
Pixels with notably negative flux ($< -2 \times 10^{-11} \ \rm cm^{-2} \, s^{-1} \, sr^{-1} \, MeV^{-1}$) in the right panel of figure \ref{fig:map_FB} are excluded from the ROI, and the negative residual map is removed from the set of the fitting templates. 
The fit result is shown in figure \ref{fig:spec_syst} (middle); the halo excess spectrum does not differ significantly from the baseline analysis.

\subsubsection{Point sources}

In the baseline analysis, point sources are subtracted using the spectral models in the 4FGL-DR4 catalog and the LAT PSF. Instead, the fit result after masking circular areas around the point sources with a radius equal to the 95\% containment angle is shown in figure \ref{fig:spec_syst} (middle).
The result is almost unchanged from the baseline analysis.
Due to the deterioration of the LAT angular resolution, the area to be masked increases toward lower photon energies, resulting in a large statistical error at the lowest bin.

\subsubsection{The GALPROP models}
\label{sec:GALPROP}

In the study of the Galactic diffuse emission, the LAT team used 128 sets of various combinations of the five GALPROP parameters \cite{Ackermann2012-CR}.
In the present study, five cases are tested in which the value of one parameter among the five is changed from the parameter set used in the baseline analysis.
The baseline GALPROP parameter set $^S$L$^Z$6$^R$30$^T$150$^C$2 means the Lorimer \cite{Lorimer2006} distribution of pulsars for the cosmic-ray source distribution, the vertical and radial boundaries of the cosmic-ray confinement region $z_h =6$ and $R_h = 30$ kpc, the spin temperature $T_S = 150$ K for the opacity of the 21 cm line, and the magnitude cut at 2 mag for the reddening $E(B-V)$ by dust as a gas tracer. 
Figure \ref{fig:spec_syst} (bottom) shows the results of five different fits in which these parameters are changed as follows:
(1) the supernova remnant (SNR) distribution \cite{Case1998} for cosmic-ray sources, (2) $z_h =$ 10 kpc, (3) $R_h =$ 20 kpc, (4) $T_S = 10^5$ K, and (5) the magnitude cut 5 mag for $E(B-V)$. 
In most cases, the fit results are almost unchanged from the baseline analysis, but relatively large changes are observed when the cosmic-ray source distribution is changed to SNR.
In this case, the halo component is positively fitted at all energy bins and is nearly zero at the lowest energy bin.
This suggests that one reason for the negative halo fits at the lowest energies in the baseline analysis may be the uncertainty of the gas component of GALPROP. 

In the LAT team paper on the GC GeV excess \cite{Ackermann2017-GCE}, the GALPROP gas component is divided into five regions of Galactocentric radius (0--1.5, 1.5--3.5, 3.5--8, 8--10, and 10--50 kpc), and the ICS component is divided into the three sub-components of the different target photon radiation fields (optical, infrared, and CMB), which are fitted as independent templates.
To check the dependence on the number of free GALPROP components, the NFW-$\rho^2$ fit is performed with the same treatment about GALPROP as ref. \cite{Ackermann2017-GCE}, and the result is shown in figure \ref{fig:spec_syst} (middle).
The halo component decreases in bins below 12 GeV from the baseline analysis, but remains almost unchanged at the 21 GeV bin, and the overall trend of the spectrum peaking at 20 GeV remains unchanged.

\subsubsection{The GIEM as the zero point}
\label{sec:GIEM}

Finally, it is examined whether the halo-like excess can also be seen when the {\it Fermi} LAT GIEM model of interstellar emission \cite{Acero2016-GIEM,Fermi-LAT2019-GIEM}, \verb|gll_iem_v07.fits|, is used as the zero point. 
Though this background model is recommended for general point source analysis, it is {\it not} recommended for studying medium or large-scale structures.
The reason is that diffuse structures such as the Fermi bubbles, the GC excess, and Loop I, for which it is difficult to create reliable templates from other wavelength data, are modeled by iteratively generating non-template patches from the fit residuals.
Any components with unknown structures or shapes may be absorbed by the patch during this process, making unbiased exploration of a new diffuse component difficult.

However, the 20 GeV halo excess found in this study may still remain as an excess from the GIEM, because the spectrum of the GIEM patch is ultimately modeled by a doubly broken power-law with breaks at 150 MeV and 3 GeV.
Since the 20 GeV halo excess is difficult to describe by a single power law above 3 GeV, the excess is expected to remain relative to the GIEM.
In the baseline analysis of the present study, the model templates are fitted independently for each energy bin without assumptions on their spectra, while the flux distribution in the sky is fixed in a model map template. 
On the other hand, the GIEM patch is created to accommodate an arbitrary sky map without using templates, while imposing restrictions on the energy spectrum.
Therefore, this test complements the baseline analysis.

This test is performed by a fit that includes the GIEM, point sources, and the isotropic background with non-zero initial values in MCMC (the original model normalization for the GIEM), as well as the NFW-$\rho^2$ model with a zero initial value.
However, with this treatment, any map component that a simple power law spectrum cannot describe will remain as an excess. 
To avoid confusion between the halo and other components, the model components used in the baseline analysis (GALPROP gas and ICS, Loop I, and the structured Fermi bubble template) are also included in the fit with zero initial values in MCMC.
The halo component spectrum obtained by this fit is shown in figure \ref{fig:spec_syst} (middle).
The halo component decreases compared to the baseline analysis, but is still statistically significant with the same peak at around 20 GeV (6.7, 7.4, and 5.1$\sigma$ at 12, 21, and 35 GeV bins, respectively).
Since a part of the halo component may have been reduced by the non-template patch in the GIEM, this result can be considered as a lower limit for the halo component.

\section{Implications for Dark Matter Annihilation}
\label{sec:DM}

\subsection{The possible origins of the halo-like excess}
\label{sec:origins}

The results up to the previous section suggest that a halo-like component physically exists, with a non-power-law spectrum peaking at around 20 GeV and the morphology close to the NFW-$\rho^2$ model.
The energy spectrum and morphology are consistent with what would be expected from dark matter annihilation, but other possibilities should also be considered.
In principle, if there is an astrophysical source with such a spectrum and morphology, it can also explain the excess. 
However, no such astrophysical object or phenomenon is known, and if it exists, its origin would be an interesting topic in astrophysics.

Another possibility is that there is an unexpected, energy-dependent change in the morphology of one of the components considered in the model calculations.
However, as shown in figure \ref{fig:map_models}, many of the components considered have morphologies that differ greatly from a halo, and it is not easy to expect them to undergo halo-like shape changes depending on photon energy.
The ICS component has the morphology closest to the spherical halo, but its shape in the GALPROP model is considerably flatter along the Galactic plane than the halo.
Looking at the fit spectra for NFW-$\rho^2$ (figure \ref{fig:specall_NFW2}), the halo component is less than 1/10 of the ICS at 4.3 GeV, but it is about the same or even more prominent than the ICS at 20 GeV.
Explaining this by an uncertainty of the ICS model requires a model change that significantly alters the shape of the ICS map.

If adding a halo component significantly alters the ICS spectrum, rendering it unrealistic as ICS, it would suggest that a change of the ICS morphology may have been confused with a fake halo component.
There is a sharp dip at the 59 GeV bin in the ICS spectrum of the NFW-$\rho^2$ fit (figure \ref{fig:specall_NFW2}), but such a dip is not observed in the 21 GeV bin, where the halo component is most prominent.
The 59 GeV dip is also seen, albeit shallow, in the ICS spectrum by the fit without any halo model (figure \ref{fig:specall_noDM}), and this is likely due to some anomaly in this energy bin.
In other energy bins, the ICS spectrum in figure \ref{fig:specall_NFW2} is power-law-like within the range of statistical errors and is not particularly strange as an ICS spectrum. Though further investigation of the possible degeneracy between the halo and ICS components would be worthwhile in future studies, the rest of this paper discusses interpretations based on dark matter annihilation.

\subsection{Fitting of the WIMP parameters}

Here, the mass $m_\chi$ and velocity-averaged annihilation cross-section $\langle \sigma \upsilon \rangle$ of the WIMPs are determined by spectral fitting, assuming that the halo excess is due to dark matter annihilation. 
The diffuse gamma-ray flux from annihilation per unit photon energy for an observer in a halo with a smooth density profile and no substructure (the NFW-$\rho^2$ case) is given by
\begin{eqnarray}
\frac{dF_\gamma}{dE_\gamma}  = \frac{\langle \sigma v \rangle}{8 \pi m_\chi^2}
\frac{dN_\gamma}{dE_\gamma} \int \rho^2 dl \ ,
\end{eqnarray}
where $dN_\gamma/dE_\gamma$ is the gamma-ray production from one annihilation event, and the integration is along the distance $l$ in the line of sight.
In this study, the fitted gamma-ray flux from the halo is normalized in the Galactic pole direction ($b = \pm 90^\circ$), and for the assumed MW halo model, the line-of-sight integral in this direction is
\begin{eqnarray}
\int_{\rm G.P.} \rho^2 dl = 8.93 \times 10^{14} \  [M_\odot^2 \ \rm kpc^{-5}] \ .
\end{eqnarray}

The WIMP parameters are also fitted for the case where subhalos dominate the annihilation signal, assuming that the subhalo number density is proportional to the smooth density profile (NFW-$\rho^1$).
The diffuse gamma-ray flux is expressed as
\begin{eqnarray}
\frac{dF_\gamma}{dE_\gamma}  =   \frac{1}{4 \pi} \frac{dN_\gamma}{dE_\gamma} \, 
\kappa \int \rho \, dl \ ,
\end{eqnarray}
where $\kappa$ is the annihilation rate per unit mass of dark matter, and for the assumed MW halo model, 
\begin{eqnarray}
\int_{\rm G.P.} \rho \, dl  = 1.49 \times 10^8 \ [M_\odot \ \rm kpc^{-2}] \ .
\end{eqnarray}
If dark matter decays rather than annihilates, $\kappa$ obtained by the fit to the data is related to the decay time scale $\tau_d = ( m_\chi \, \kappa)^{-1}$. The parameter $\kappa$ can also be linked to $\langle \sigma v \rangle$ in the case of annihilation by the boost factor $B$, which is widely used in the literature. 
This is defined as the total annihilation luminosity from all subhalos contained in a halo being boosted by a factor of $B$, compared with that of a smooth dark matter distribution.
The smallest subhalos expected from the structure formation theory are much smaller than the resolution of numerical simulations, and hence the uncertainty in $B$ is large.
Estimates of hundreds or even thousands have been reported for $B$, but more recent studies suggest $B \sim$ 5--30 \cite{Cirelli2024,SanchezConde2014,Moline2017}.
The combination $B \, \langle \sigma v \rangle$ can then be related to $\kappa$ as
\begin{eqnarray}
B \, \langle \sigma v \rangle = 2 m_\chi^2 \, \kappa \, 
\left[ \frac{\int \rho^2 \, r^2 \, dr}{\int \rho \, r^2 \, dr} \right]^{-1} \ , 
\end{eqnarray}
where the density integration along the Galactocentric radius $r$ for the MW halo model here is
\begin{eqnarray}
\frac{\int \rho^2 \, r^2 \, dr}{\int \rho \, r^2 \, dr} &=& 0.163 \, \rho_s \ .
\end{eqnarray}

\begin{figure}[tbp]
\centering 
\includegraphics[width=1.0\textwidth]{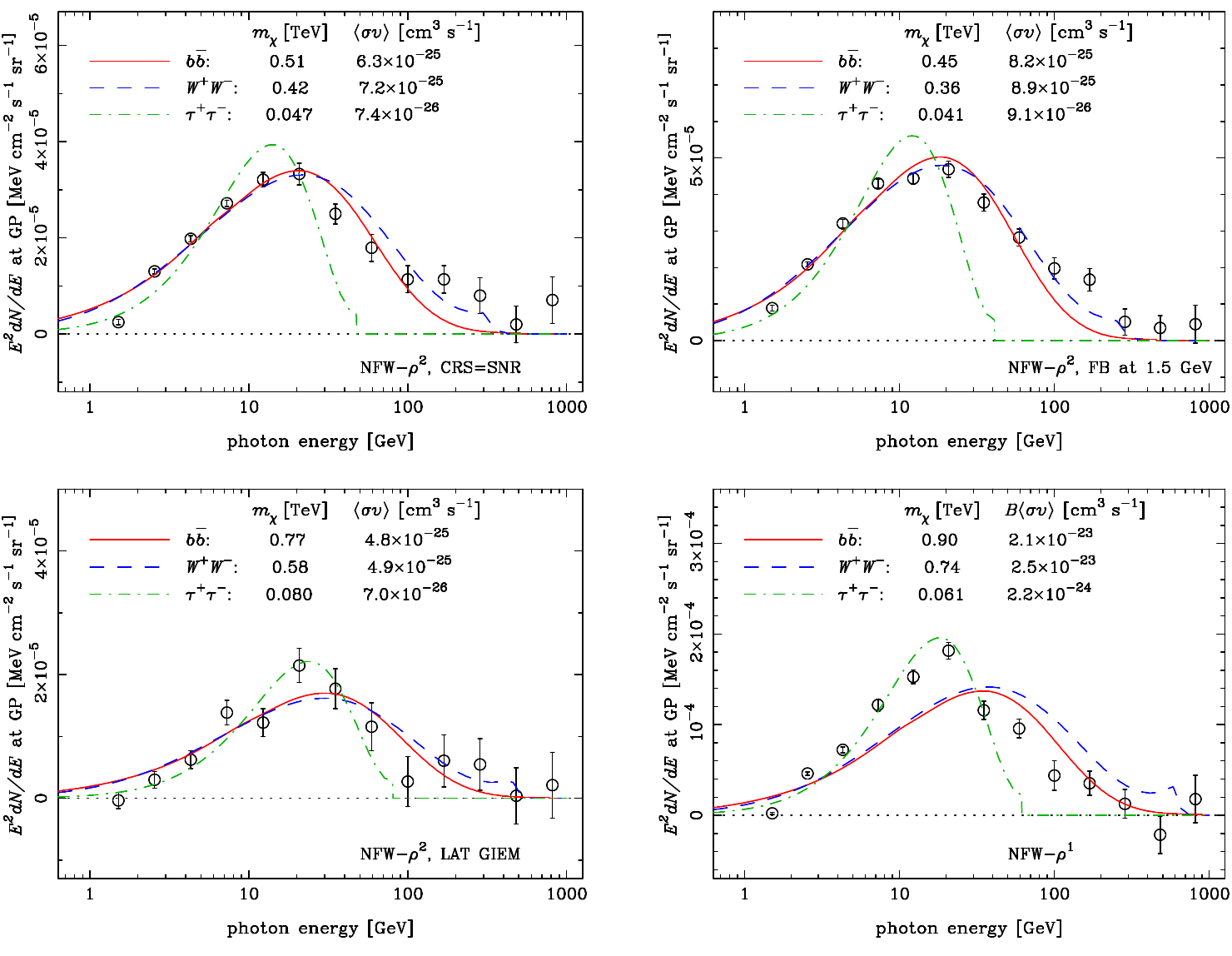}
\caption{\label{fig:spec_dmfit} 
Fits of dark matter annihilation spectra to the halo-like excess (the flux to the Galactic poles is shown). 
The fitted WIMP parameters $m_\chi$ and $\langle \sigma \upsilon \rangle$ (for NFW-$\rho^2$) or  $B \, \langle \sigma \upsilon \rangle$ (for NFW-$\rho^1$) are indicated in each panel, for the three annihilation channels of $b\bar{b}$, $W^+W^-$, and $\tau^+\tau^-$. 
For the NFW-$\rho^2$ fit, three halo-excess spectra by different analyses are shown: (1) the SNR distribution for cosmic-ray sources, (2) the Fermi bubble template created at 1.5 GeV, and (3) the halo excess relative to the GIEM background model. Error bars are statistical 1$\sigma$ estimated by MCMC. 
}
\end{figure}

An annihilation of dark matter is generally expected to produce a pair of a particle and its antiparticle in the Standard Model of particle physics, like $b\bar{b}$.
Such pairs ultimately generate numerous particles through cascading, which can be calculated using the Standard Model.
Following the literature, we present three representative cases of $b\bar{b}$, $W^+W^-$, and $\tau^+\tau^-$ as the dominant annihilation channel, and $dN_\gamma/dE_\gamma$ for these are obtained from the PPPC 4 DM ID package (Release 6.0) \cite{Cirelli2011,Cirelli2011-E,Ciafaloni2011}.

Fits for the four cases (three for NFW-$\rho^2$, and one for NFW-$\rho^1$) are shown in figure \ref{fig:spec_dmfit}, where the best-fit WIMP parameters are also shown. 
As photon energy decreases, statistical errors in the data become smaller, and hence spectral fitting tends to be influenced by the lowest energy bins.
However, as we have already seen, the systematic error is much larger than the statistical error in these bins.
Therefore, the WIMP parameters obtained here should be considered as rough estimates based on the peak position and normalization of the halo excess.
For this reason, the statistical errors of the WIMP parameters are not reported.
The NFW-$\rho^2$ halo flux of the baseline analysis is negative in the lowest energy bin, and the WIMP parameter fit is strongly affected by this unphysical negative value and its small statistical error.
Therefore, from among the alternative analyses relative to the NFW-$\rho^2$ baseline, three are selected in which the halo flux of the lowest energy bin is not strongly negative: 
(1) the SNR distribution for cosmic-ray sources in GALPROP, (2) the Fermi bubble template created at 1.5 GeV, and (3) the GIEM used as the reference in the halo excess search. 
The case (2) results in the highest halo component flux (and therefore the largest cross section), while the case (3) results in the lowest.
For the reason stated in section \ref{sec:GIEM}, the cross section of the case (3) should be considered as a lower limit.

In the NFW-$\rho^2$ fits, $m_\chi \sim$ 0.4--0.8 TeV and $\langle \sigma \upsilon \rangle \sim$  (5--9)$\times 10^{-25} \ \rm cm^3 \, s^{-1}$ are favored for the $b\bar{b}$ and $W^+W^-$ channels, and 40--80 GeV and  (7--9)$\times 10^{-26} \ \rm cm^3 \, s^{-1}$ for the $\tau^+\tau^-$ channel. 
The cross sections favored in the NFW-$\rho^1$ fit are $B \, \langle \sigma \upsilon \rangle \sim 2 \times 10^{-23} \ \rm cm^3 \, s^{-1}$  ($b\bar{b}$ and $W^+W^-$) and $2 \times 10^{-24} \ \rm cm^3 \, s^{-1}$ ($\tau^+\tau^-$). 
These correspond to the decay timescale $\tau_d \sim 2 \times 10^{27}$ and $1 \times 10^{27}$ s, respectively, when dark matter decays rather than annihilates.
In the fits for the $\tau^+\tau^-$ channel, $m_\chi$ is likely biased toward a lower value because of the small statistical errors in low-energy bins. 
The halo excess spectrum seems to fit the $b\bar{b}$ and $W^+W^-$ channels better than $\tau^+\tau^-$ due to their broader annihilation spectra, though the halo excess spectrum contains significant systematic uncertainties, particularly in the low-energy region.

\subsection{Comparison with other dark matter searches}

\subsubsection{Relationship with the GC GeV excess}

Dark matter annihilation is one of the leading candidates for the origin of the GC GeV excess.
The 20 GeV halo excess found in this study has the peak photon energy that is significantly higher than the GC excess, and the preferred density profile ($\rho \propto r^{-1}$ or shallower) is flatter than that for the GC excess ($\rho \propto r^{-1.25}$). 
Therefore, the 20 GeV halo excess and the GC excess are likely to have different origins, and hence, if the halo excess is of dark matter origin, the GC excess must be explained by another origin.
Many hypotheses have been proposed for the GC excess, and millisecond pulsars, in particular, are considered to be a hypothesis that is as promising as dark matter \cite{Murgia2020}.

It should be noted that the halo excess has a flux similar to that of the GC excess at $|b| > 10^\circ$ and around 2 GeV, where the GC excess peaks (see figures \ref{fig:specall_FB} and \ref{fig:specall_NFW2}).
If past studies have confused part of the halo excess with the extended tail of the GC excess, the true GC excess may be more centrally concentrated than $\rho \propto r^{-1.25}$.
Extension of the GC excess to more than 10$^\circ$ has been argued to support the dark matter origin \cite{Calore2015-JCAP,Calore2015-PRD,Daylan2016}, but if the true GC excess is more centrally concentrated, it may be more consistent with stellar population origins, such as millisecond pulsars.

\subsubsection{Comparison with previous constraints and theoretical expectations}

The annihilation cross section estimated from the 20 GeV halo excess ($\langle \sigma \upsilon \rangle \sim 6 \times 10^{-25} \ \rm cm^3 \, s^{-1}$ with $m_\chi \sim 500$ GeV for the $b\bar{b}$ channel) is marginally consistent with the upper limits obtained by earlier studies on the MW halo region \cite{Ackermann2012-MWhalo,Mazziotta2012,GomezVargas2013,Tavakoli2014,Chang2018}. 
The LAT team's analysis using two years of the LAT data \cite{Ackermann2012-MWhalo} reported $\langle \sigma \upsilon \rangle \lesssim 1 \times 10^{-24} \ \rm cm^3 s^{-1}$ without background modeling, for their standard choice of the dark matter density at the solar neighborhood, $\rho_\odot = 0.43 \ \rm GeV \, cm^{-3}$. It should also be noted that the ROI of the LAT team's analysis ($5^\circ \le |b| \le 15^\circ$, $|l| \le 80^\circ$) is much narrower in the Galactic latitude than that of this study.

The cross section required for the halo excess is larger than the upper limits by dwarf spheroidal galaxies, which are currently considered to be the strongest \cite{Ackermann2015-dSph,Albert2017,Acciari2020,McDaniel2024}. 
For example, it is several times larger than the 95\% containment region of the constraints obtained in a recent study \cite{McDaniel2024}.
However, this tension does not immediately rule out the dark matter interpretation of the 20 GeV halo excess.
Compared to dwarf galaxies, whose dark matter mass can be more reliably constrained by observations, the density distribution of the MW halo is highly uncertain (\cite{Cirelli2024}, see also section \ref{sec:MW_uncertainty} below).
The dark matter content in dwarf galaxies also remains subject to uncertainties regarding methodology and observed data \cite{Ando2020,Yang2025}.
In fact, the dark matter interpretation of the GC GeV excess is also in a similar degree of tension with dwarf galaxy constraints, but for the same reason, the dark matter interpretation is not considered to be excluded \cite{Murgia2020,McDaniel2024}.

The annihilation cross section determines the era of WIMP decoupling in the early universe, and is therefore related to the relic amount of dark matter in the present universe. 
The $b\bar{b}$-channel cross section estimated from the 20 GeV halo excess is more than an order of magnitude larger than the canonical thermal relic cross section to explain the present dark matter density, $\langle \sigma \upsilon \rangle \sim$ (2--3)$\times 10^{-26} \ \rm cm^3 \, s^{-1}$ \cite{Roszkowski2018,Cirelli2024}. 
However, considering the uncertainty and degrees of freedom in particle physics theory, there are various possibilities that the actual annihilation cross section differs from this by orders of magnitude.
Considering also the uncertainty in the MW halo density profile, the dark matter interpretation of the 20 GeV halo excess is not inconsistent with theoretical expectations.

\subsubsection{Uncertainties in the MW dark matter density distribution}
\label{sec:MW_uncertainty}

The uncertainties in the density profile of the MW halo can be described in more detail as follows.
Since the mass of baryons exceeds that of dark matter in the inner region of the MW ($r \lesssim$ 10 kpc) \cite{Sofue2009,Prantzos2011,Nesti2013,Portail2017}, we can only extrapolate from the dark matter density at the solar system ($\rho_\odot$) assuming a halo density profile.
In addition to the uncertainty of the density profile, the estimate for $\rho_\odot$ has an uncertainty of a factor of about two centered around $0.4 \ \rm GeV \, cm^{-3}$ \cite{Cirelli2024}.
In the case of a smooth density distribution ($\epsilon_\gamma \propto \rho^2$), a threefold uncertainty in the dark matter density at 10$^\circ$--20$^\circ$ from the GC changes the annihilation flux by nearly an order of magnitude.

There are several other analytic functions proposed for the halo density profile, in addition to NFW, and the Einasto profile \cite{Einasto1965}, which is known to fit numerical simulations as well as NFW, is particularly interesting for the halo excess. 
When $\rho_\odot$ is fixed, this profile predicts a higher density than NFW at around 10$^\circ$ ($r \sim 1.4$ kpc) from the GC, and a shallower profile than NFW near the GC (see figure 2.3 of \cite{Cirelli2024}), which are in agreement with the trends suggested by the 20 GeV halo excess. 
Some reports imply that a more core-like density profile than NFW better matches observational data for the MW halo \cite{Nesti2013,Portail2017}.

In addition to the uncertainty in the fitting formulae for the mean halo density profile in numerical simulations, individual halo profiles also vary depending on their formation and merger history.
Furthermore, in the central region of a halo, dark matter density profile may be affected by complicated baryon dynamics, and this effect is not yet fully understood \cite{Tollet2016,Katz2016}.  
Finally, the effect of subhalos or any substructures in the MW halo should not be overlooked.
Since substructures are more easily destroyed in the central region \cite{Kuhlen2008}, the signal may be boosted with an effective density profile shallower than NFW, in a direction consistent with the trends of the halo excess.

\subsubsection{The extragalactic gamma-ray background radiation}

If the MW halo emits annihilation gamma-rays, the extragalactic gamma-ray background radiation (EGB) should include a contribution from annihilations in all dark halos in the universe.
Theoretical predictions suggest that the EGB flux by annihilation is comparable to that from the MW halo, with the peak photon energy redshifted to about half \cite{Fermi-LAT2015-DM-EGB}.
The observed EGB in the 1--1000 GeV range is thought to contain various astronomical components, such as active galactic nuclei and star-forming galaxies \cite{Fornasa2015}, making it difficult to identify the dark matter component unambiguously.
A recent study suggests that a significant contribution to the EGB from annihilation with the WIMP parameters of the 20 GeV halo excess is not ruled out \cite{Cholis2024}.

The EGB from star-forming galaxies is well understood in terms of physical processes, and relatively reliable theoretical models have been developed \cite{Fornasa2015,Chen2025}. 
These models contribute significantly to the {\it Fermi} LAT data below 10 GeV, but their predicted flux decreases rapidly above 10 GeV due to intergalactic absorption. 
On the other hand, the EGB data extend approximately as a power law beyond 10 GeV, and a new component may be appearing in this region.
Annihilation gamma rays from the MW halo and extragalactic halos, whose spectral peak is around 10--20 GeV, may be a candidate for such a component.
In particular, the emission from the MW halo is expected to be relatively stronger at higher photon energies because it is not subject to intergalactic absorption, and some of it may have been confused with the isotropic background radiation.
In fact, this was the original motivation behind this study to search for signs of dark matter annihilation from the MW halo.

\section{Summary and Discussion}
\label{sec:summary}

In this study, 15 years of the {\it Fermi} LAT data have been used to search for annihilation gamma rays from dark matter in the halo region of the MW ($|l| \le 60^\circ$, $|b| \le 60^\circ$), in the photon energy range of 1.5--810 GeV. 
Following the standard analyses by the LAT team, the gamma-ray intensity map in the ROI is modeled considering the following components: point sources, the interstellar gas and ICS components of GALPROP, isotropic background radiation, Loop I, and the Fermi bubbles. 
Then it is investigated whether a halo-like component is necessary in addition to these model components, in the ROI excluding the Galactic plane region ($|b| < 10^\circ$).
To focus on the smooth structure of the halo scale, photon counts for the likelihood are calculated in each analysis cell of size $\Delta l_c = \Delta b_c = 10^\circ$.

The model fitting analysis favors the presence of a halo component with a spectral peak around 20 GeV at high statistical significance, in contrast to the GC GeV excess that peaks at 2--3 GeV.
The flux of the halo component is consistent with zero at energies below 2 GeV and above 200 GeV.
The radial profile fits well with the NFW-$\rho^2$ model (the NFW profile and gamma-ray emissivity per unit volume for the smooth distribution, $\epsilon_\gamma \propto \rho^2$), rather than $\epsilon_\gamma \propto \rho^{2.5}$ (GC GeV excess) or $\epsilon_\gamma \propto \rho^{1}$ (subhalos or decaying dark matter). 
To avoid exceeding the data when extrapolating the halo component to the GC, a profile that is slightly shallower than NFW-$\rho^2$ is preferred, especially in the central region.
Looking at the maps of the halo model plus fit residual, the shapes are nearly spherically symmetric in all energy bins where the halo component is detected, indicating that the halo model is a good description of the observed map. 

Various types of systematic uncertainties are examined, including the GALPROP model parameters and the morphology of the Fermi bubble template. 
Due to the strong diffuse emission from cosmic-ray interactions, the systematic uncertainty of the halo component flux becomes larger at lower photon energies. 
However, the trend that the halo component rises from 1 GeV, peaks at around 20 GeV, and becomes almost undetectable above 200 GeV remains unchanged.
In particular, it is noteworthy that the halo-like excess is detected with a similar spectrum, even when the LAT standard background model (GIEM) is used as the zero point. 
Since the GIEM includes a non-template patch created from the fit residuals, this suggests that the halo excess is not an artifact dependent on the model map templates used in the baseline analysis. 

While the possibility of astrophysical explanation cannot be ruled out, the characteristics of this ``Fermi halo'' are consistent with what would be expected for dark matter annihilation signals.
The spectrum of the halo excess can be fitted with the most representative annihilation channel into $b\bar{b}$, and in that case, the mass and annihilation cross section of the WIMPs are estimated to be $m_\chi \sim$ 0.5--0.8 TeV and $\langle \sigma \upsilon \rangle \sim$ (5--8)$\times 10^{-25} \ \rm cm^3 \, s^{-1}$, respectively. 
This cross section is in tension with the upper limits derived from dwarf spheroidal galaxies, and is more than an order of magnitude larger than the canonical thermal relic cross section. 
However, considering various uncertainties, particularly those in the density profile of the MW halo, the dark matter interpretation of the Fermi halo is compatible with these.

However, the signal from the MW halo alone does not constitute the definitive proof of dark matter annihilation.
Detection of annihilation signals from other objects or regions with consistent WIMP parameters will be crucial for the final confirmation.
Gamma-ray observations of dwarf galaxies in the MW halo are fascinating from this perspective.
Given the tension between the MW halo excess and the dwarf galaxy constraints, a modest improvement in the sensitivity of future observations should eventually lead to the detection of annihilation gamma rays.
It is noteworthy that interesting excesses have already been observed in several dwarf galaxies, even if they have not been definitively detected \cite{McDaniel2024}.
The largest excess has been reported in Reticulum II by several studies (typically $\sim 2$--$3\sigma$) \cite{DrlicaWagner2015,GeringerSameth2015,Hooper2015,DiMauro2021,McDaniel2024}, and the WIMP mass estimated by the latest study \cite{McDaniel2024} is interestingly similar to that for the MW halo excess found by the present study. 

Another important avenue is to search for line gamma-ray emission.
Line emissions at the photon energy equal to the WIMP mass (0.3--0.8 TeV for the $b\bar{b}$ channel) are expected, although the intensity will depend on the branching ratio of annihilation into two photons (or perhaps $Z\gamma$).
In this energy range, ground-based atmospheric Cherenkov telescopes are more sensitive than the {\it Fermi} LAT, and further significant improvements in sensitivity are expected shortly by the Cherenkov Telescope Array Observatory (CTAO) \cite{Abe2024-CTALine}.

Dark matter searches using neutrinos are also intriguing. 
Though the current neutrino constraints \cite{Cirelli2024,Abe2020-SK,Albert2020-ANTARES,Abbasi2023-IceCube} are 1--2 orders of magnitude (depending on the annihilation channels) weaker than the WIMP parameters suggested in this work, future improvements in the sensitivity of neutrino detectors may enable independent verification.
Since the gamma-ray to neutrino flux ratio can be calculated using the Standard Model of particle physics, the neutrino flux from the Fermi halo can be predicted without uncertainty about the dark matter density profile.


\acknowledgments

The author thanks Tsunefumi Mizuno for the useful discussion and advice. 
The author was supported by the JSPS/MEXT KAKENHI Grant Number
18K03692.


\bibliography{totani_dm2025} 

\providecommand{\href}[2]{#2}\begingroup\raggedright\begin{thebibliography}{10}

\bibitem{Zwicky1933}
F.~Zwicky, \emph{Die rotverschiebung von extragalaktischen nebeln},
  {\emph{Helvetica Physica Acta} {\bfseries 6} (1933) 110}.

\bibitem{Bertone2018}
G.~Bertone and D.~Hooper, \emph{History of dark matter},
  \href{https://doi.org/10.1103/revmodphys.90.045002}{\emph{Reviews of Modern
  Physics} {\bfseries 90} (2018) 045002}.

\bibitem{Arbey2021}
A.~Arbey and F.~Mahmoudi, \emph{Dark matter and the early universe: A review},
  \href{https://doi.org/10.1016/j.ppnp.2021.103865}{\emph{Progress in Particle
  and Nuclear Physics} {\bfseries 119} (2021) 103865}.

\bibitem{Cirelli2024}
M.~{Cirelli}, A.~{Strumia} and J.~{Zupan}, \emph{{Dark Matter}},
  \href{https://doi.org/10.48550/arXiv.2406.01705}{\emph{arXiv e-prints} (2024)
  arXiv:2406.01705} [\href{https://arxiv.org/abs/2406.01705}{{\ttfamily
  2406.01705}}].

\bibitem{Arcadi2018}
G.~{Arcadi}, M.~{Dutra}, P.~{Ghosh}, M.~{Lindner}, Y.~{Mambrini}, M.~{Pierre}
  et~al., \emph{{The waning of the WIMP? A review of models, searches, and
  constraints}},
  \href{https://doi.org/10.1140/epjc/s10052-018-5662-y}{\emph{European Physical
  Journal C} {\bfseries 78} (2018) 203}
  [\href{https://arxiv.org/abs/1703.07364}{{\ttfamily 1703.07364}}].

\bibitem{Roszkowski2018}
L.~Roszkowski, E.M.~Sessolo and S.~Trojanowski, \emph{Wimp dark matter
  candidates and searches—current status and future prospects},
  \href{https://doi.org/10.1088/1361-6633/aab913}{\emph{Reports on Progress in
  Physics} {\bfseries 81} (2018) 066201}.

\bibitem{Ackermann2017-GCE}
M.~Ackermann, M.~Ajello, A.~Albert, W.B.~Atwood, L.~Baldini, J.~Ballet et~al.,
  \emph{The fermi galactic center gev excess and implications for dark matter},
  \href{https://doi.org/10.3847/1538-4357/aa6cab}{\emph{The Astrophysical
  Journal} {\bfseries 840} (2017) 43}.

\bibitem{Murgia2020}
S.~Murgia, \emph{The fermi-lat galactic center excess: Evidence of annihilating
  dark matter?},
  \href{https://doi.org/10.1146/annurev-nucl-101916-123029}{\emph{Annual Review
  of Nuclear and Particle Science} {\bfseries 70} (2020) 455}.

\bibitem{Abdalla2022}
H.~{Abdalla}, F.~{Aharonian}, F.A.~{Benkhali}, E.O.~{Ang{\"u}ner}, C.~{Armand},
  H.~{Ashkar} et~al., \emph{{Search for Dark Matter Annihilation Signals in the
  H.E.S.S. Inner Galaxy Survey}},
  \href{https://doi.org/10.1103/PhysRevLett.129.111101}{\emph{Physical Review
  Letters} {\bfseries 129} (2022) 111101}
  [\href{https://arxiv.org/abs/2207.10471}{{\ttfamily 2207.10471}}].

\bibitem{Buckley2015}
M.R.~Buckley, E.~Charles, J.M.~Gaskins, A.M.~Brooks, A.~Drlica-Wagner,
  P.~Martin et~al., \emph{Search for gamma-ray emission from dark matter
  annihilation in the large magellanic cloud with the fermi large area
  telescope}, \href{https://doi.org/10.1103/physrevd.91.102001}{\emph{Physical
  Review D} {\bfseries 91} (2015) 102001}.

\bibitem{Caputo2016}
R.~Caputo, M.~Buckley, P.~Martin, E.~Charles, A.~Brooks, A.~Drlica-Wagner
  et~al., \emph{Search for gamma-ray emission from dark matter annihilation in
  the small magellanic cloud with the fermi large area telescope},
  \href{https://doi.org/10.1103/physrevd.93.062004}{\emph{Physical Review D}
  {\bfseries 93} (2016) 062004}.

\bibitem{Ackermann2015-dSph}
M.~Ackermann, A.~Albert, B.~Anderson, W.B.~Atwood, L.~Baldini, G.~Barbiellini
  et~al., \emph{Searching for dark matter annihilation from milky way dwarf
  spheroidal galaxies with six years of fermi large area telescope data},
  \href{https://doi.org/10.1103/physrevlett.115.231301}{\emph{Physical Review
  Letters} {\bfseries 115} (2015) 231301}.

\bibitem{Albert2017}
A.~Albert, B.~Anderson, K.~Bechtol, A.~Drlica-Wagner, M.~Meyer,
  M.~Sanchez-Conde et~al., \emph{Searching for dark matter annihilation in
  recently discovered milky way satellites with fermi-lat},
  \href{https://doi.org/10.3847/1538-4357/834/2/110}{\emph{The Astrophysical
  Journal} {\bfseries 834} (2017) 110}.

\bibitem{Acciari2020}
V.~Acciari, S.~Ansoldi, L.~Antonelli, A.~Arbet~Engels, D.~Baack, A.~Babic
  et~al., \emph{A search for dark matter in triangulum ii with the magic
  telescopes}, \href{https://doi.org/10.1016/j.dark.2020.100529}{\emph{Physics
  of the Dark Universe} {\bfseries 28} (2020) 100529}.

\bibitem{McDaniel2024}
A.~McDaniel, M.~Ajello, C.M.~Karwin, M.~Di~Mauro, A.~Drlica-Wagner and
  M.A.~Sanchez-Conde, \emph{Legacy analysis of dark matter annihilation from
  the milky way dwarf spheroidal galaxies with 14 years of fermi-lat data},
  \href{https://doi.org/10.1103/physrevd.109.063024}{\emph{Physical Review D}
  {\bfseries 109} (2024) 063024}.

\bibitem{Ackermann2012-subhalo}
M.~Ackermann, A.~Albert, L.~Baldini, J.~Ballet, G.~Barbiellini, D.~Bastieri
  et~al., \emph{Search for dark matter satellites using fermi-lat},
  \href{https://doi.org/10.1088/0004-637x/747/2/121}{\emph{The Astrophysical
  Journal} {\bfseries 747} (2012) 121}.

\bibitem{Coronado2022}
J.~Coronado-Blazquez, M.A.~Sanchez-Conde, J.~Perez-Romero and
  A.~Aguirre-Santaella, \emph{Spatial extension of dark subhalos as seen by
  fermi-lat and the implications for wimp constraints},
  \href{https://doi.org/10.1103/physrevd.105.083006}{\emph{Physical Review D}
  {\bfseries 105} (2022) 083006}.

\bibitem{Albert2018}
A.~Albert, R.~Alfaro, C.~Alvarez, J.~lavarez, R.~Arceo, J.~Arteaga-Velazquez
  et~al., \emph{Search for dark matter gamma-ray emission from the andromeda
  galaxy with the high-altitude water cherenkov observatory},
  \href{https://doi.org/10.1088/1475-7516/2018/06/043}{\emph{Journal of
  Cosmology and Astroparticle Physics} {\bfseries 2018} (2018) 043}.

\bibitem{Karwin2021}
C.M.~Karwin, S.~Murgia, I.V.~Moskalenko, S.P.~Fillingham, A.-K.~Burns and
  M.~Fieg, \emph{Dark matter interpretation of the fermi-lat observations
  toward the outer halo of m31},
  \href{https://doi.org/10.1103/physrevd.103.023027}{\emph{Physical Review D}
  {\bfseries 103} (2021) 023027}.

\bibitem{Arlen2012}
T.~Arlen, T.~Aune, M.~Beilicke, W.~Benbow, A.~Bouvier, J.H.~Buckley et~al.,
  \emph{Constraints on cosmic rays, magnetic fields, and dark matter from
  gamma-ray observations of the coma cluster of galaxies with veritas and
  fermi}, \href{https://doi.org/10.1088/0004-637x/757/2/123}{\emph{The
  Astrophysical Journal} {\bfseries 757} (2012) 123}.

\bibitem{Ackermann2015-cl}
M.~Ackermann, M.~Ajello, A.~Albert, W.B.~Atwood, L.~Baldini, G.~Barbiellini
  et~al., \emph{Search for extended gamma-ray emission from the virgo galaxy
  cluster with fermi-lat},
  \href{https://doi.org/10.1088/0004-637x/812/2/159}{\emph{The Astrophysical
  Journal} {\bfseries 812} (2015) 159}.

\bibitem{Lisanti2018}
M.~Lisanti, S.~Mishra-Sharma, N.L.~Rodd and B.R.~Safdi, \emph{Search for dark
  matter annihilation in galaxy groups},
  \href{https://doi.org/10.1103/physrevlett.120.101101}{\emph{Physical Review
  Letters} {\bfseries 120} (2018) 101101}.

\bibitem{DiMauro2023}
M.~Di~Mauro, J.~Perez-Romero, M.A.~Sanchez-Conde and N.~Fornengo,
  \emph{Constraining the dark matter contribution of $\gamma$ rays in clusters
  of galaxies using fermi-lat data},
  \href{https://doi.org/10.1103/physrevd.107.083030}{\emph{Physical Review D}
  {\bfseries 107} (2023) 083030}.

\bibitem{Fermi-LAT2015-DM-EGB}
{The Fermi LAT collaboration}, \emph{Limits on dark matter annihilation signals
  from the fermi lat 4-year measurement of the isotropic gamma-ray background},
  \href{https://doi.org/10.1088/1475-7516/2015/09/008}{\emph{Journal of
  Cosmology and Astroparticle Physics} {\bfseries 2015} (2015) 008}.

\bibitem{Fornasa2015}
M.~Fornasa and M.A.~Sanchez-Conde, \emph{The nature of the diffuse gamma-ray
  background},
  \href{https://doi.org/10.1016/j.physrep.2015.09.002}{\emph{Physics Reports}
  {\bfseries 598} (2015) 1}.

\bibitem{Fornasa2016}
M.~Fornasa, A.~Cuoco, J.~Zavala, J.M.~Gaskins, M.A.~Sanchez-Conde,
  G.~Gomez-Vargas et~al., \emph{Angular power spectrum of the diffuse gamma-ray
  emission as measured by the fermi large area telescope and constraints on its
  dark matter interpretation},
  \href{https://doi.org/10.1103/physrevd.94.123005}{\emph{Physical Review D}
  {\bfseries 94} (2016) }.

\bibitem{Cholis2024}
I.~Cholis and I.~Krommydas, \emph{Scrutinizing the isotropic gamma-ray
  background in search of dark matter},
  \href{https://doi.org/10.1103/physrevd.110.103032}{\emph{Physical Review D}
  {\bfseries 110} (2024) 103032}.

\bibitem{Ackermann2012-MWhalo}
M.~Ackermann, M.~Ajello, W.B.~Atwood, L.~Baldini, G.~Barbiellini, D.~Bastieri
  et~al., \emph{Constraints on the galactic halo dark matter from fermi-lat
  diffuse measurements},
  \href{https://doi.org/10.1088/0004-637x/761/2/91}{\emph{The Astrophysical
  Journal} {\bfseries 761} (2012) 91}.

\bibitem{Mazziotta2012}
M.~Mazziotta, F.~Loparco, F.~de~Palma and N.~Giglietto, \emph{A
  model-independent analysis of the fermi large area telescope gamma-ray data
  from the milky way dwarf galaxies and halo to constrain dark matter
  scenarios},
  \href{https://doi.org/10.1016/j.astropartphys.2012.07.005}{\emph{Astroparticle
  Physics} {\bfseries 37} (2012) 26}.

\bibitem{GomezVargas2013}
G.A.~Gomez-Vargas, M.A.~Sanchez-Conde, J.-H.~Huh, M.~Peiro, F.~Prada,
  A.~Morselli et~al., \emph{Constraints on wimp annihilation for contracted
  dark matter in the inner galaxy with the fermi-lat},
  \href{https://doi.org/10.1088/1475-7516/2013/10/029}{\emph{Journal of
  Cosmology and Astroparticle Physics} {\bfseries 2013} (2013) 029}.

\bibitem{Tavakoli2014}
M.~Tavakoli, I.~Cholis, C.~Evoli and P.~Ullio, \emph{Constraints on dark matter
  annihilations from diffuse gamma-ray emission in the galaxy},
  \href{https://doi.org/10.1088/1475-7516/2014/01/017}{\emph{Journal of
  Cosmology and Astroparticle Physics} {\bfseries 2014} (2014) 017}.

\bibitem{Ackermann2015-line}
M.~Ackermann, M.~Ajello, A.~Albert, B.~Anderson, W.~Atwood, L.~Baldini et~al.,
  \emph{Updated search for spectral lines from galactic dark matter
  interactions with pass 8 data from the fermi large area telescope},
  \href{https://doi.org/10.1103/physrevd.91.122002}{\emph{Physical Review D}
  {\bfseries 91} (2015) 122002}.

\bibitem{Chang2018}
L.J.~Chang, M.~Lisanti and S.~Mishra-Sharma, \emph{Search for dark matter
  annihilation in the milky way halo},
  \href{https://doi.org/10.1103/physrevd.98.123004}{\emph{Physical Review D}
  {\bfseries 98} (2018) 123004}.

\bibitem{Albert2023}
A.~Albert, R.~Alfaro, C.~Alvarez, J.~Arteaga-Velazquez, D.~Avila~Rojas,
  H.~Ayala~Solares et~al., \emph{An optimized search for dark matter in the
  galactic halo with hawc},
  \href{https://doi.org/10.1088/1475-7516/2023/12/038}{\emph{Journal of
  Cosmology and Astroparticle Physics} {\bfseries 2023} (2023) 038}.

\bibitem{Goodenough2009}
L.~{Goodenough} and D.~{Hooper}, \emph{{Possible Evidence For Dark Matter
  Annihilation In The Inner Milky Way From The Fermi Gamma Ray Space
  Telescope}}, \href{https://doi.org/10.48550/arXiv.0910.2998}{\emph{arXiv
  e-prints} (2009) arXiv:0910.2998}
  [\href{https://arxiv.org/abs/0910.2998}{{\ttfamily 0910.2998}}].

\bibitem{Vitale2009}
V.~{Vitale} and A.~{Morselli}, \emph{{Indirect Search for Dark Matter from the
  center of the Milky Way with the Fermi-Large Area Telescope}},
  \href{https://doi.org/10.48550/arXiv.0912.3828}{\emph{arXiv e-prints} (2009)
  arXiv:0912.3828} [\href{https://arxiv.org/abs/0912.3828}{{\ttfamily
  0912.3828}}].

\bibitem{Dobler2010}
G.~Dobler, D.P.~Finkbeiner, I.~Cholis, T.~Slatyer and N.~Weiner, \emph{The
  fermi haze: A gamma-ray counterpart to the microwave haze},
  \href{https://doi.org/10.1088/0004-637x/717/2/825}{\emph{The Astrophysical
  Journal} {\bfseries 717} (2010) 825}.

\bibitem{Su2010}
M.~Su, T.R.~Slatyer and D.P.~Finkbeiner, \emph{Giant gamma-ray bubbles from
  fermi-lat: Active galactic nucleus activity or bipolar galactic wind?},
  \href{https://doi.org/10.1088/0004-637x/724/2/1044}{\emph{The Astrophysical
  Journal} {\bfseries 724} (2010) 1044}.

\bibitem{Ackermann2014-FB}
M.~Ackermann, A.~Albert, W.B.~Atwood, L.~Baldini, J.~Ballet, G.~Barbiellini
  et~al., \emph{The spectrum and morphology of the fermi bubbles},
  \href{https://doi.org/10.1088/0004-637x/793/1/64}{\emph{The Astrophysical
  Journal} {\bfseries 793} (2014) 64}.

\bibitem{Navarro1997}
J.F.~Navarro, C.S.~Frenk and S.D.M.~White, \emph{A universal density profile
  from hierarchical clustering},
  \href{https://doi.org/10.1086/304888}{\emph{The Astrophysical Journal}
  {\bfseries 490} (1997) 493}.

\bibitem{Ajello2021}
M.~Ajello, W.B.~Atwood, M.~Axelsson, R.~Bagagli, M.~Bagni, L.~Baldini et~al.,
  \emph{Fermi large area telescope performance after 10 years of operation},
  \href{https://doi.org/10.3847/1538-4365/ac0ceb}{\emph{The Astrophysical
  Journal Supplement Series} {\bfseries 256} (2021) 12}.

\bibitem{Abdollahi2020}
S.~Abdollahi, F.~Acero, M.~Ackermann, M.~Ajello, W.B.~Atwood, M.~Axelsson
  et~al., \emph{Fermi large area telescope fourth source catalog},
  \href{https://doi.org/10.3847/1538-4365/ab6bcb}{\emph{The Astrophysical
  Journal Supplement Series} {\bfseries 247} (2020) 33}.

\bibitem{Abdollahi2022}
S.~Abdollahi, F.~Acero, L.~Baldini, J.~Ballet, D.~Bastieri, R.~Bellazzini
  et~al., \emph{Incremental fermi large area telescope fourth source catalog},
  \href{https://doi.org/10.3847/1538-4365/ac6751}{\emph{The Astrophysical
  Journal Supplement Series} {\bfseries 260} (2022) 53}.

\bibitem{Strong2007}
A.W.~{Strong}, I.V.~{Moskalenko} and V.S.~{Ptuskin}, \emph{{Cosmic-Ray
  Propagation and Interactions in the Galaxy}},
  \href{https://doi.org/10.1146/annurev.nucl.57.090506.123011}{\emph{Annual
  Review of Nuclear and Particle Science} {\bfseries 57} (2007) 285}
  [\href{https://arxiv.org/abs/astro-ph/0701517}{{\ttfamily
  astro-ph/0701517}}].

\bibitem{Vladimirov2011}
A.~Vladimirov, S.~Digel, G.~Johannesson, P.~Michelson, I.~Moskalenko, P.~Nolan
  et~al., \emph{Galprop webrun: An internet-based service for calculating
  galactic cosmic ray propagation and associated photon emissions},
  \href{https://doi.org/10.1016/j.cpc.2011.01.017}{\emph{Computer Physics
  Communications} {\bfseries 182} (2011) 1156}.

\bibitem{Porter2017}
T.A.~Porter, G.~Johannesson and I.V.~Moskalenko, \emph{High-energy gamma rays
  from the milky way: Three-dimensional spatial models for the cosmic-ray and
  radiation field densities in the interstellar medium},
  \href{https://doi.org/10.3847/1538-4357/aa844d}{\emph{The Astrophysical
  Journal} {\bfseries 846} (2017) 67}.

\bibitem{Ackermann2012-CR}
M.~Ackermann, M.~Ajello, W.B.~Atwood, L.~Baldini, J.~Ballet, G.~Barbiellini
  et~al., \emph{Fermi-lat observations of the diffuse $\gamma$-ray emission:
  Implications for cosmic rays and the interstellar medium},
  \href{https://doi.org/10.1088/0004-637x/750/1/3}{\emph{The Astrophysical
  Journal} {\bfseries 750} (2012) 3}.

\bibitem{Acero2016-GIEM}
F.~Acero, M.~Ackermann, M.~Ajello, A.~Albert, L.~Baldini, J.~Ballet et~al.,
  \emph{Development of the model of galactic interstellar emission for standard
  point-source analysis of fermi large area telescope data},
  \href{https://doi.org/10.3847/0067-0049/223/2/26}{\emph{The Astrophysical
  Journal Supplement Series} {\bfseries 223} (2016) 26}.

\bibitem{Fermi-LAT2019-GIEM}
{The Fermi LAT collaboration}, \emph{Galactic interstellar emission model for
  the 4fgl catalog analysis}, {\emph{A pdf document available at the {\it
  Fermi} website:
  \url{https://fermi.gsfc.nasa.gov/ssc/data/analysis/software/aux/4fgl/Galactic_Diffuse_Emission_Model_for_the_4FGL_Catalog_Analysis.pdf}}
  (2019) }.

\bibitem{Large1962}
M.I.~Large, M.J.S.~Quigley and C.G.T.~Haslam, \emph{A new feature of the radio
  sky}, \href{https://doi.org/10.1093/mnras/124.5.405}{\emph{Monthly Notices of
  the Royal Astronomical Society} {\bfseries 124} (1962) 405}.

\bibitem{Wolleben2007}
M.~Wolleben, \emph{A new model for the loop i (north polar spur) region},
  \href{https://doi.org/10.1086/518711}{\emph{The Astrophysical Journal}
  {\bfseries 664} (2007) 349}.

\bibitem{Kataoka2013}
J.~Kataoka, M.~Tahara, T.~Totani, Y.~Sofue, U.~Stawarz, Y.~Takahashi et~al.,
  \emph{Suzaku observations of the diffuse x-ray emission across the fermi
  bubbles’ edges},
  \href{https://doi.org/10.1088/0004-637x/779/1/57}{\emph{The Astrophysical
  Journal} {\bfseries 779} (2013) 57}.

\bibitem{Casandjian2009}
J.-M.~{Casandjian} and I.~{Grenier}, \emph{{High Energy Gamma-Ray Emission from
  the Loop I region}},
  \href{https://doi.org/10.48550/arXiv.0912.3478}{\emph{arXiv e-prints} (2009)
  arXiv:0912.3478} [\href{https://arxiv.org/abs/0912.3478}{{\ttfamily
  0912.3478}}].

\bibitem{Kuhlen2008}
M.~Kuhlen, J.~Diemand and P.~Madau, \emph{The dark matter annihilation signal
  from galactic substructure: Predictions for glast},
  \href{https://doi.org/10.1086/590337}{\emph{The Astrophysical Journal}
  {\bfseries 686} (2008) 262}.

\bibitem{Pieri2011}
L.~Pieri, J.~Lavalle, G.~Bertone and E.~Branchini, \emph{Implications of
  high-resolution simulations on indirect dark matter searches},
  \href{https://doi.org/10.1103/physrevd.83.023518}{\emph{Physical Review D}
  {\bfseries 83} (2011) 023518}.

\bibitem{Ackermann2015-EGRB}
M.~Ackermann, M.~Ajello, A.~Albert, W.B.~Atwood, L.~Baldini, J.~Ballet et~al.,
  \emph{The spectrum of isotropic diffuse gamma-ray emission between 100 mev
  and 820 gev}, \href{https://doi.org/10.1088/0004-637x/799/1/86}{\emph{The
  Astrophysical Journal} {\bfseries 799} (2015) 86}.

\bibitem{Lorimer2006}
D.R.~Lorimer, A.J.~Faulkner, A.G.~Lyne, R.N.~Manchester, M.~Kramer,
  M.A.~McLaughlin et~al., \emph{The parkes multibeam pulsar survey -- vi.
  discovery and timing of 142 pulsars and a galactic population analysis},
  \href{https://doi.org/10.1111/j.1365-2966.2006.10887.x}{\emph{Monthly Notices
  of the Royal Astronomical Society} {\bfseries 372} (2006) 777}.

\bibitem{Case1998}
G.L.~Case and D.~Bhattacharya, \emph{A new $\sigma$-$d$ relation and its
  application to the galactic supernova remnant distribution},
  \href{https://doi.org/10.1086/306089}{\emph{The Astrophysical Journal}
  {\bfseries 504} (1998) 761}.

\bibitem{SanchezConde2014}
M.A.~S\'anchez-Conde and F.~Prada, \emph{The flattening of the
  concentration-mass relation towards low halo masses and its implications for
  the annihilation signal boost},
  \href{https://doi.org/10.1093/mnras/stu1014}{\emph{Monthly Notices of the
  Royal Astronomical Society} {\bfseries 442} (2014) 2271}.

\bibitem{Moline2017}
{\'A}.~{Molin{\'e}}, M.A.~{S{\'a}nchez-Conde}, S.~{Palomares-Ruiz} and
  F.~{Prada}, \emph{{Characterization of subhalo structural properties and
  implications for dark matter annihilation signals}},
  \href{https://doi.org/10.1093/mnras/stx026}{\emph{Monthly Notices of the
  Royal Astronomical Society} {\bfseries 466} (2017) 4974}
  [\href{https://arxiv.org/abs/1603.04057}{{\ttfamily 1603.04057}}].

\bibitem{Cirelli2011}
M.~Cirelli, G.~Corcella, A.~Hektor, G.~H\"{u}tsi, M.~Kadastik, P.~Panci et~al.,
  \emph{Pppc 4 dm id: a poor particle physicist cookbook for dark matter
  indirect detection},
  \href{https://doi.org/10.1088/1475-7516/2011/03/051}{\emph{Journal of
  Cosmology and Astroparticle Physics} {\bfseries 2011} (2011) 051}.

\bibitem{Cirelli2011-E}
M.~Cirelli, G.~Corcella, A.~Hektor, G.~H\"{u}tsi, M.~Kadastik, P.~Panci et~al.,
  \emph{Erratum: Pppc 4 dm id: a poor particle physicist cookbook for dark
  matter indirect detection},
  \href{https://doi.org/10.1088/1475-7516/2012/10/e01}{\emph{Journal of
  Cosmology and Astroparticle Physics} {\bfseries 2012} (2012) E01}.

\bibitem{Ciafaloni2011}
P.~Ciafaloni, D.~Comelli, A.~Riotto, F.~Sala, A.~Strumia and A.~Urbano,
  \emph{Weak corrections are relevant for dark matter indirect detection},
  \href{https://doi.org/10.1088/1475-7516/2011/03/019}{\emph{Journal of
  Cosmology and Astroparticle Physics} {\bfseries 2011} (2011) 019}.

\bibitem{Calore2015-JCAP}
F.~Calore, I.~Cholis and C.~Weniger, \emph{Background model systematics for the
  fermi gev excess},
  \href{https://doi.org/10.1088/1475-7516/2015/03/038}{\emph{Journal of
  Cosmology and Astroparticle Physics} {\bfseries 2015} (2015) 038}.

\bibitem{Calore2015-PRD}
F.~Calore, I.~Cholis, C.~McCabe and C.~Weniger, \emph{A tale of tails: Dark
  matter interpretations of the fermi gev excess in light of background model
  systematics},
  \href{https://doi.org/10.1103/physrevd.91.063003}{\emph{Physical Review D}
  {\bfseries 91} (2015) 063003}.

\bibitem{Daylan2016}
T.~Daylan, D.P.~Finkbeiner, D.~Hooper, T.~Linden, S.K.~Portillo, N.L.~Rodd
  et~al., \emph{The characterization of the gamma-ray signal from the central
  milky way: A case for annihilating dark matter},
  \href{https://doi.org/10.1016/j.dark.2015.12.005}{\emph{Physics of the Dark
  Universe} {\bfseries 12} (2016) 1}.

\bibitem{Ando2020}
S.~Ando, A.~Geringer-Sameth, N.~Hiroshima, S.~Hoof, R.~Trotta and M.G.~Walker,
  \emph{Structure formation models weaken limits on wimp dark matter from dwarf
  spheroidal galaxies},
  \href{https://doi.org/10.1103/physrevd.102.061302}{\emph{Physical Review D}
  {\bfseries 102} (2020) 061302}.

\bibitem{Yang2025}
H.~Yang, W.~Wang, L.~Zhu, T.S.~Li, S.E.~Koposov, J.~Han et~al., \emph{The dark
  matter content of milky way dwarf spheroidal galaxies: Draco, sextans and
  ursa minor}, {\emph{arXiv e-prints} (2025) arXiv:2507.02284}
  [\href{https://arxiv.org/abs/2507.02284}{{\ttfamily 2507.02284}}].

\bibitem{Sofue2009}
Y.~Sofue, M.~Honma and T.~Omodaka, \emph{Unified rotation curve of the galaxy
  --- decomposition into de vaucouleurs bulge, disk, dark halo, and the 9-kpc
  rotation dip ---},
  \href{https://doi.org/10.1093/pasj/61.2.227}{\emph{Publications of the
  Astronomical Society of Japan} {\bfseries 61} (2009) 227}.

\bibitem{Prantzos2011}
N.~Prantzos, C.~Boehm, A.M.~Bykov, R.~Diehl, K.~Ferriere, N.~Guessoum et~al.,
  \emph{The 511 kev emission from positron annihilation in the galaxy},
  \href{https://doi.org/10.1103/revmodphys.83.1001}{\emph{Reviews of Modern
  Physics} {\bfseries 83} (2011) 1001}.

\bibitem{Nesti2013}
F.~Nesti and P.~Salucci, \emph{The dark matter halo of the milky way, ad 2013},
  \href{https://doi.org/10.1088/1475-7516/2013/07/016}{\emph{Journal of
  Cosmology and Astroparticle Physics} {\bfseries 2013} (2013) 016}.

\bibitem{Portail2017}
M.~Portail, O.~Gerhard, C.~Wegg and M.~Ness, \emph{Dynamical modelling of the
  galactic bulge and bar: the milky way’s pattern speed, stellar and dark
  matter mass distribution},
  \href{https://doi.org/10.1093/mnras/stw2819}{\emph{Monthly Notices of the
  Royal Astronomical Society} {\bfseries 465} (2017) 1621}.

\bibitem{Einasto1965}
J.~Einasto, \emph{On the construction of a composite model for the galaxy and
  on the determination of the system of galactic parameters}, {\emph{Trudy
  Astrofizicheskogo Instituta Alma-Ata} {\bfseries 5} (1965) 87}.

\bibitem{Tollet2016}
E.~Tollet, A.V.~Maccio, A.A.~Dutton, G.S.~Stinson, L.~Wang, C.~Penzo et~al.,
  \emph{Nihao - iv: core creation and destruction in dark matter density
  profiles across cosmic time},
  \href{https://doi.org/10.1093/mnras/stv2856}{\emph{Monthly Notices of the
  Royal Astronomical Society} {\bfseries 456} (2016) 3542}.

\bibitem{Katz2016}
H.~Katz, F.~Lelli, S.S.~McGaugh, A.~Di~Cintio, C.B.~Brook and J.M.~Schombert,
  \emph{Testing feedback-modified dark matter haloes with galaxy rotation
  curves: estimation of halo parameters and consistency with $\lambda$cdm
  scaling relations},
  \href{https://doi.org/10.1093/mnras/stw3101}{\emph{Monthly Notices of the
  Royal Astronomical Society} {\bfseries 466} (2016) 1648}.

\bibitem{Chen2025}
J.~Chen and T.~Totani, \emph{The diffuse extragalactic $\gamma$-ray background
  radiation: star-forming galaxies are not the dominant component},
  \href{https://doi.org/10.1093/mnras/staf928}{\emph{Monthly Notices of the
  Royal Astronomical Society} {\bfseries 540} (2025) 3221}.

\bibitem{DrlicaWagner2015}
A.~Drlica-Wagner, A.~Albert, K.~Bechtol, M.~Wood, L.~Strigari, M.~Sanchez-Conde
  et~al., \emph{Search for gamma-ray emission from des dwarf spheroidal galaxy
  candidates with fermi-lat data},
  \href{https://doi.org/10.1088/2041-8205/809/1/l4}{\emph{The Astrophysical
  Journal} {\bfseries 809} (2015) L4}.

\bibitem{GeringerSameth2015}
A.~Geringer-Sameth, M.G.~Walker, S.M.~Koushiappas, S.E.~Koposov, V.~Belokurov,
  G.~Torrealba et~al., \emph{Indication of gamma-ray emission from the newly
  discovered dwarf galaxy reticulum ii},
  \href{https://doi.org/10.1103/physrevlett.115.081101}{\emph{Physical Review
  Letters} {\bfseries 115} (2015) 081101}.

\bibitem{Hooper2015}
D.~Hooper and T.~Linden, \emph{On the gamma-ray emission from reticulum ii and
  other dwarf galaxies},
  \href{https://doi.org/10.1088/1475-7516/2015/09/016}{\emph{Journal of
  Cosmology and Astroparticle Physics} {\bfseries 2015} (2015) 016}.

\bibitem{DiMauro2021}
M.~Di~Mauro and M.W.~Winkler, \emph{Multimessenger constraints on the dark
  matter interpretation of the fermi-lat galactic center excess},
  \href{https://doi.org/10.1103/physrevd.103.123005}{\emph{Physical Review D}
  {\bfseries 103} (2021) 123005}.

\bibitem{Abe2024-CTALine}
S.~Abe, J.~Abhir, A.~Abhishek, F.~Acero, A.~Acharyya, R.~Adam et~al.,
  \emph{Dark matter line searches with the cherenkov telescope array},
  \href{https://doi.org/10.1088/1475-7516/2024/07/047}{\emph{Journal of
  Cosmology and Astroparticle Physics} {\bfseries 2024} (2024) 047}.

\bibitem{Abe2020-SK}
K.~Abe, C.~Bronner, Y.~Haga, Y.~Hayato, M.~Ikeda, S.~Imaizumi et~al.,
  \emph{Indirect search for dark matter from the galactic center and halo with
  the super-kamiokande detector},
  \href{https://doi.org/10.1103/physrevd.102.072002}{\emph{Physical Review D}
  {\bfseries 102} (2020) 072002}.

\bibitem{Albert2020-ANTARES}
A.~Albert, M.~Andre, M.~Anghinolfi, G.~Anton, M.~Ardid, J.-J.~Aubert et~al.,
  \emph{Search for dark matter towards the galactic centre with 11 years of
  antares data},
  \href{https://doi.org/10.1016/j.physletb.2020.135439}{\emph{Physics Letters
  B} {\bfseries 805} (2020) 135439}.

\bibitem{Abbasi2023-IceCube}
R.~Abbasi, M.~Ackermann, J.~Adams, S.~Agarwalla, J.~Aguilar, M.~Ahlers et~al.,
  \emph{Search for neutrino lines from dark matter annihilation and decay with
  icecube}, \href{https://doi.org/10.1103/physrevd.108.102004}{\emph{Physical
  Review D} {\bfseries 108} (2023) 102004}.

\end{thebibliography}\endgroup
\bibliographystyle{JHEP}

\end{document}